\documentclass{article}

 \usepackage[preprint,nonatbib]{neurips_2023}

\usepackage[utf8]{inputenc} 
\usepackage[T1]{fontenc}    
\usepackage[breaklinks=true,colorlinks,bookmarks=false]{hyperref}
\usepackage{url}            
\usepackage{booktabs}       
\usepackage{amsfonts}       
\usepackage{nicefrac}       
\usepackage{microtype}      
\usepackage{xcolor}         
\usepackage{amsmath}
\usepackage{graphicx} 
\usepackage{multirow}
\usepackage{adjustbox}
\usepackage{graphicx}
\usepackage{subcaption}
\usepackage{float}

\newcommand{\difficulty}[1]{\textcolor{pink}{[\difficulty:#1]}}

\title{Evaluating the Robustness of Text-to-image \\Diffusion Models against Real-world Attacks

}
\usepackage{caption}
\author{%
  Hongcheng Gao$^{1, 3}$, Hao Zhang$^4$, Yinpeng Dong$^2$\thanks{Corresponding authors.}\;, Zhijie Deng$^1$\footnotemark[1] \\
  $^1$Shanghai Jiao Tong University\quad $^2$Tsinghua University \\ 
  $^3$Chongqing University \quad $^4$University of California San Diego\\
  \texttt{gaohongcheng2000@gmail.com; sjtu.haozhang@gmail.com;}\\ \texttt{dongyinpeng@mail.tsinghua.edu.cn; zhijied@sjtu.edu.cn} \\
}

\begin{document}

\maketitle

\begin{abstract}
Text-to-image (T2I) diffusion models (DMs) have shown promise in generating high-quality images from textual descriptions. The real-world applications of these models require particular attention to their safety and fidelity, but this has not been sufficiently explored. 
One fundamental question is whether existing T2I DMs are robust against variations over input texts. 
To answer it, this work provides the first robustness evaluation of T2I DMs against \emph{real-world} attacks. 
Unlike prior studies that focus on malicious attacks involving apocryphal alterations to the input texts, we consider an attack space spanned by realistic errors (e.g., typo, glyph, phonetic) that humans can make, to ensure semantic consistency. Given the inherent randomness of the generation process, we develop novel distribution-based attack objectives to mislead T2I DMs.
We perform attacks in a black-box manner without any knowledge of the model. Extensive experiments demonstrate the effectiveness of our method for attacking popular T2I DMs and simultaneously reveal their non-trivial robustness issues. 
Moreover, we provide an in-depth analysis of our method to show that it is not designed to attack the text encoder in T2I DMs solely.
\end{abstract}

\section{Introduction}
Diffusion models (DMs)~\cite{sohl2015deep, NEURIPS2020_4c5bcfec,song2020denoising} have demonstrated remarkable success in generating images and shown promise in diverse applications, including super-resolution~\cite{saharia2022image}, image inpainting
~\cite{lugmayr2022repaint}, text-to-image synthesis~\cite{DBLP:conf/cvpr/RombachBLEO22,ramesh2022hierarchical}, video generation~\cite{ho2022imagen,ho2022video}, etc.
A typical DM employs a forward process that gradually diffuses the data distribution towards a noise distribution and a reverse process that recovers the data through step-by-step denoising. 
Among the applications, text-to-image (T2I) generation has received significant attention and witnessed the development of large models such as GLIDE~\cite{nichol2022glide}, Imagen~\cite{saharia2022photorealistic}, DALL-E 2~\cite{ramesh2022hierarchical}, Stable Diffusion~\cite{DBLP:conf/cvpr/RombachBLEO22}, VQ-Diffusion~\cite{DBLP:conf/cvpr/GuCBWZCYG22}, etc. These models typically proceed by conditioning the reverse process on the embeddings of textual descriptions obtained from certain text encoders. 
Their ability to generate high-quality images from textual descriptions can significantly simplify the creation of game scenarios, book illustrations, organization logos, and more. 

However, the ability of T2I DMs to generate high-quality content also raises ethical concerns about their potential misuse. 
In particular, they may be used to produce fake imagery of existing individuals for misinformation 
~\cite{carlini2020evading}, or yield visual content deemed offensive or harmful~\cite{chen2023pathway}. 
These concerns have been used to justify the decision to limit access to large T2I DMs, as well as moderate their use according to content policies implemented in prompt filters. 
To summarize, the real-world adoption of T2I DMs raises demands for serious consideration of their safety and fidelity. 


Evaluating the robustness of T2I DMs is a fundamental problem in this regard, often achieved by adversarial attacks. Some initial studies in this field have demonstrated that manipulating the input text by creating meaningless or distorted custom words~\cite{milliere2022adversarial} or phrases~\cite{maus2023adversarial}, or adding irrelevant distractions~\cite{zhuang2023pilot} can lead to significant bias in the semantics of the images generated by T2I DMs.
However, it should be noted that these works primarily focus on malicious attacks, which often introduce substantial changes to the text and may rarely occur in real-world scenarios. To bridge this gap, we suggest shifting the focus from intentional attacks to everyday errors such as typos, grammar mistakes, or vague expressions, as suggested by related works in natural language processing~\cite{li2018textbugger, eger-benz-2020-hero,Eger-2019-viper,le2022perturbations}, to thoroughly evaluate the robustness of models that interact with humans in \emph{pratical} use. 
This work provides the first evaluation of the robustness of T2I DMs against \emph{real-world} attacks. 
As discussed, we consider an attack space spanned by realistic errors that humans can make to ensure semantic consistency, including typos, glyphs, and phonetics. 
To tackle the inherent uncertainty in the generation process of DMs, we develop novel distribution-based attack objectives to mislead T2I DMs. 
We perform attacks in a black-box manner using greedy search to avoid assumptions about the model. 
Technically, our attack algorithm first identifies the keywords based on the words' marginal influence on the generation distribution and then applies elaborate character-level replacements. 



\begin{figure}
    \centering
        \includegraphics[page=5,width=\textwidth]{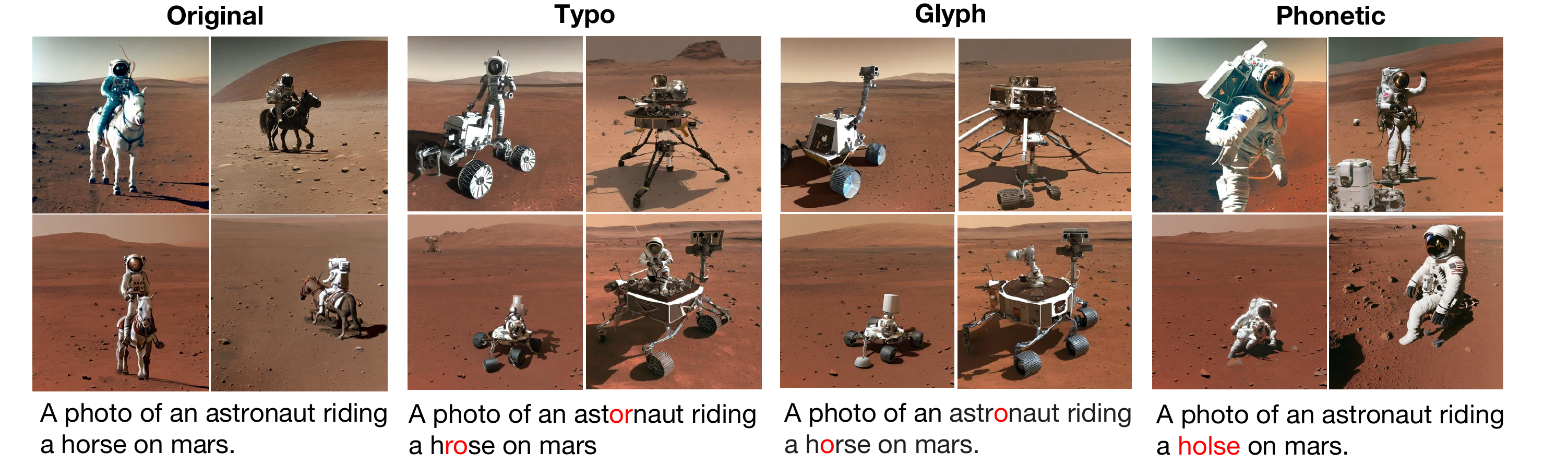}
  \caption{An illustration of our attack method against Stable Diffusion~\cite{DBLP:conf/cvpr/RombachBLEO22} based on three attack rules (detailed in Section~\ref{sec:rules}). Adversarially modified content is highlighted in \color[HTML]{FE0000} red \color[HTML]{000000}.}
  \vspace{-10pt}
  \label{image_example}
\end{figure}


We perform extensive empirical evaluations on datasets of artificial prompts and image captions. 
We first conduct a set of diagnostic experiments to prioritize the different variants originating from the distribution-oriented attack goal, which also reflects the vulnerability of existing T2I DMs. 
We then provide an interesting discussion on the target of attacking DMs: only the text encoder in them or the whole diffusion process? 
Finally, we perform attacks against T2I DMs in real-world settings and observe high success rates, even in the case that the perturbation rates and query times are low. 



    

\section{Related Work}
\textbf{Diffusion models}
(DMs)~\cite{sohl2015deep, NEURIPS2020_4c5bcfec,song2020denoising} have achieved great success in image synthesis recently. 
In the diffusion process, the data distribution is diffused to an isotropic Gaussian by continually adding Gaussian noises. 
The reverse process recovers the original input from a Gaussian noise by denoising. 
DMs have been widely applied to text-to-image (T2I) generation. 
GLIDE~\cite{nichol2022glide} first achieves this by integrating the text feature into transformer blocks in the denoising process. 
Subsequently, increasing effort is devoted to this field to improve the performance of T2I generation, with DALL-E~\cite{ramesh2021zero}, Cogview~\cite{ding2021cogview}, Make-A-Scene~\cite{gafni2022make}, Stable Diffusion~\cite{DBLP:conf/cvpr/RombachBLEO22}, and Imagen~\cite{saharia2022photorealistic} as popular examples. 
A prevalent strategy nowadays is to perform denoising in the feature space while introducing the text condition by cross-attention mechanisms~\cite{tang2022daam}.
However, textual conditions can not provide the synthesis results with more structural guidance. 
To remediate this, there are many other kinds of DMs conditioning on factors beyond text descriptions, such as PITI~\cite{wang2022pretraining}, ControlNet~ \cite{zhang2023adding} and 
Sketch-Guided models~\cite{voynov2022sketch}. 

\textbf{Adversarial attack}~\cite{DBLP:journals/corr/SzegedyZSBEGF13,zhang2020adversarial} typically deceive DNNs by integrating carefully-crafted tiny perturbations into input data. Based on how an adversary interacts with the victim model, adversarial attacks can be categorized into white-box attacks~\cite{zhang2022investigating,meng2020geometry,xu-etal-2022-exploring} (with full access to the victim model based on which attacks are generated) and black-box attacks~\cite{Zhang_2022_CVPR,he2021model} (with access only to the victim model’s input and output). Adversarial attacks on text can also be categorized in terms of the level of granularity of the perturbation. Character-level~\cite{formento2023using,eger2019text} attacks modify individual characters in words to force the tokenizer to process multiple unrelated embeddings instead of the original, resulting in decreased performance. Word-level~\cite{lee2022query,li2021contextualized} attacks employ a search algorithm to locate useful perturbing embeddings or operations that are clustered close to the candidate attack word’s embedding given a similarity constraint (e.g., the Universal Sentence Encoder~\cite{cer2018universal}). 
Sentence-level~\cite{wang2020t3,han2020adversarial} attack refers to making changes to sentence structures in order to prevent the model from correctly predicting the outcome. Multi-level~\cite{gupta2021synthesizing,wallace2019universal} attacks combine multiple types of perturbations, making the attack cumulative. Recent studies~\cite{milliere2022adversarial,maus2023adversarial,zhuang2023pilot} have explored the over-sensitivity of T2I DMs to prompt perturbations in the text domain with malicious word synthesis, phrase synthesis, and adding distraction. \cite{zhuang2023pilot} also reveals the vulnerability of T2I models and attributes it to the weak robustness of the used text encoders.

\section{Methodology}


This section provides a detailed description of our approach to real-world adversarial attacks of T2I DMs. 
We briefly outline the problem formulation before delving into the design of attack objective functions and then describe how to perform optimization in a black-box manner. 


\subsection{Problem Formulation}
A T2I DM that accepts text inputs $c$ and generates images $x$ essentially characterizes the conditional distribution $p_\theta(x | c)$ with $\theta$ as model parameters. 
To evaluate the robustness of modern DMs so as to govern their behaviors when adopted in the wild, 
we opt to attack the input text, i.e., finding a text $c'$ which keeps close to the original text $c$ but can lead to a significantly biased generated distribution. 
Such an attack is meaningful in the sense of encompassing real-world perturbations such as typos, glyphs, and phonetics.
Concretely, 
the optimization problem is formulated as follows:
\begin{equation}
\label{eq:1}
    \max_{c'} \mathcal{D}(p_\theta(x | c') \Vert p_\theta(x | c)), \quad \text{s.t.}\; d(c, c') \leq \epsilon,
\end{equation}
where $\mathcal{D}$ denotes a divergence measure between two distributions, $d(c, c')$ measures the distance between two texts, and $\epsilon$ indicates the perturbation budget. 

The main challenge of attack lies in that we cannot write down the exact formulation of $p_\theta(x | c)$ and $p_\theta(x | c')$ of DMs but get only a few i.i.d. samples $\{\bar{x}_1, \dots, \bar{x}_N\}$ and $\{x_1, \dots, x_N\}$\ 
from them, where $\bar{x}_i$ is an image generated with the original text $c$ while $x_i$ is generated with the modified text $c'$. 


\subsection{Attack Objective}
Here, we develop four instantiations of the distribution-based attack objective defined in Eq.~\eqref{eq:1}.

\subsubsection{MMD Distance} 
As validated by the community~\cite{tolstikhin2016minimax,dziugaite2015training}, maximum mean discrepancy (MMD) distance is frequently used for distinguishing two distributions given finite samples. 
Formally, assuming access to a kernel function $\kappa$, the square of MMD distance is typically defined as:
\begin{equation}
\small
    \mathcal{D}_{\text{MMD}^2}(p_\theta(x | c') \Vert p_\theta(x | c)) \approx \frac{1}{N^2} \sum_{i=1}^N\sum_{j=1}^N \kappa(x_i, x_j) - \frac{2}{N^2} \sum_{i=1}^N\sum_{j=1}^N \kappa(x_i, \bar{x}_j) + C, 
\end{equation}
where $C$ refers to a constant agnostic to $c'$. 
The feature maps associated with the kernel should be able to help construct useful statistics of the sample set such that MMD can compare distributions. 
In the case that $x$ represents generated images, a valid choice is a deep kernel built upon a pre-trained NN-based image encoder $h$ (e.g., a ViT trained by the objective of MAE~\cite{he2022masked} or CLIP~\cite{radford2021learning}). 
In practice, we specify the kernel with a simple cosine form {\small $\kappa(x, x') := {h(x)^\top h(x')}/{\Vert h(x) \Vert \Vert h(x') \Vert}$} given that $h$'s outputs usually locate in a well-suited Euclidean space.

\subsubsection{KL Divergence}
Considering that text also provides crucial information in the attack process, we will incorporate text information to consider the joint distribution of images and texts. Due to the excellent ability of CLIP to represent both image and text information while preserving their relationships, we have chosen to use CLIP as the model for encoding images and texts. Assume access to a pre-trained $\phi$-parameterized CLIP model comprised of an image encoder $h_{\phi}$ and a text encoder $g_{\phi}$ and assume the output features to be L$^2$-normalized. 
It can provide a third-party characterization of the joint distribution between the image $x$ and the text $c$ for guiding attack.
Note that $h_{\phi}(x)^\top g_{\phi}(c)$ measures the likelihood of the coexistence of image $x$ and text $c$, thus from a probabilistic viewpoint, we can think of $e_\phi(x, c) := \alpha h_{\phi}(x)^\top g_{\phi}(c)$, where $\alpha$ is some constant scaling factor, as $\log p_\phi(x, c)$. 
Under the mild assumption that $ p_\phi(x | c)$ approximates $p_\theta(x|c)$, we instantiate the measure $\mathcal{D}$ in Eq.~(\ref{eq:1}) with KL divergence and derive the following maximization objective (details are deferred to Appendix):
\begin{equation}
\label{eq:kl}
\begin{aligned}
\centering
    &\mathcal{D}_\text{KL}(p_\theta(x | c') \Vert p_\theta(x | c)) \approx \mathbb{E}_{p_\theta(x | c')} [-{ e_\phi(x, c)}] + \mathbb{E}_{p_\theta(x | c')} [\log {p_\theta(x | c')}]  + C,
\end{aligned}
\end{equation}
where $C$ denotes a constant agnostic to $c'$. 
The first term corresponds to generating images containing semantics contradictory to text $c$ and can be easily computed by Monte Carlo (MC) estimation. 
The second term is negative entropy, so the maximization of it means reducing generation diversity. 
Whereas, in practice, the entropy of distribution over high-dimensional images cannot be trivially estimated given a few samples. 
To address this issue, 
we replace $\mathbb{E}_{p_\theta(x | c')} [\log {p_\theta(x | c')}]$ with a lower bound $\mathbb{E}_{p_\theta(x | c')} [\log {q(x)}]$ for any probability distribution $q$, due to that $\mathcal{D}_\text{KL}(p_\theta(x | c') \Vert q(x)) = \mathbb{E}_{p_\theta(x | c')} [\log {p_\theta(x | c')} - \log {q(x)}] \geq 0$. 
In practice, we can only acquire distributions associated with the CLIP model, so we primarily explore the following two strategies. 

\noindent \textbf{Strategy 1.} $\log q(x) := \log p_\phi(x, c') = e_\phi(x, c')$. 
Combining with Eq.~(\ref{eq:kl}), there is ($C$ is omitted):
\begin{equation}
\small
\begin{aligned}
\label{eq:kl1}
    \mathcal{D}_\text{KL}(p_\theta(x | c') \Vert p_\phi(x | c)) & \geq \mathbb{E}_{p_\theta(x | c')} [-{ e_\phi(x, c)} + { e_\phi(x, c')}]  \approx \alpha \Big[\frac{1}{N} \sum_{i=1}^N h_{\phi}(x_i)\Big]^\top   \Big(g_{\phi}(c') - g_{\phi}(c)\Big).
\end{aligned}
\end{equation}
The adversarial text $c'$ would affect both the generated images $x_i$ and the text embeddings $g_{\phi}(c')$. 
Therefore, it is likely that by maximizing the resulting term in Eq.~(\ref{eq:kl1}) w.r.t. $c'$, the text encoder of the CLIP model is attacked (i.e., $g_{\phi}(c') - g_{\phi}(c)$ is pushed to align with the average image embedding), which deviates from our goal of delivering a biased generation distribution. 

\noindent \textbf{Strategy 2.} $\log q(x) := \log p_\phi(x) =\mathrm{L}_{\hat{c} \in \mathcal{C}}( e_\phi(x, \hat{c})) - \log |\mathcal{C}|$ where $\mathrm{L}$ is the log-sum-exp operator and $\mathcal{C}$ denotes the set of all possible text inputs. 
Likewise, there is (we omit constants):
\begin{equation}
\small
\begin{aligned}
\label{eq:kl2}
    \mathcal{D}_\text{KL}(p_\theta(x | c') \Vert p_\phi(x | c)) & \geq \mathbb{E}_{p_\theta(x | c')} [-{ e_\phi(x, c)} + \mathrm{L}_{\hat{c} \in \mathcal{C}}( e_\phi(x, \hat{c}))]  \approx \frac{1}{N} \sum_{i=1}^N \big[\mathrm{L}_{\hat{c} \in \mathcal{C}}( e_\phi(x_i, \hat{c})) - e_\phi(x_i, c)\big].
\end{aligned}
\end{equation}
As shown, the first term pushes the generated images toward the high-energy regions, and the second term hinders the generated images from containing semantics about $c$. 
To reduce the computational overhead, we draw a set of commonly used texts and pre-compute their text embeddings via CLIP before attacking. 
Then, during attacking, we only need to send the embeddings of generated images to a linear transformation followed by an $\mathrm{L}$ operator to get an estimation of the first term of Eq.~(\ref{eq:kl2}). 

\subsubsection{Two-sample Test} 
In essence, distinguishing $p_\theta(x | c')$ and $p_\theta(x | c)$ by finite observations corresponds to a two-sample test (2ST) in statistics, and the aforementioned MMD distance is a test statistic that  gains particular attention in the machine learning community. 
Based on this point, we are then interested in building a general framework that can embrace existing off-the-shelf two-sample test tools for attacking T2I DMs.
This can considerably enrich the modeling space. 
Basically, we define a unified test statistic in the following formula
\begin{equation}
\label{eq:4}
   \hat{t}\Big(\{\varphi(x_i)\}_{i=1}^N, \{\varphi(\bar{x}_i)\}_{i=1}^N\Big).
\end{equation}
Roughly speaking, we will reject the null hypothesis $p_\theta(x | c') = p_\theta(x | c)$ when the statistic is large to a certain extent.
The function $\hat{t}$ in the above equation is customized by off-the-shelf two-sample test tools such as KS test, t-test, etc.
Considering the behavior of these tools may quickly deteriorate as the dimension increases~\cite{gretton2012kernel}, we introduce a projector $\varphi$ to produce one-dimensional representations of images $x$. 
As a result, $\varphi$ implicitly determines the direction of our attack. 
For example, if we define $\varphi$ as a measurement of image quality in terms of FID~\cite{heusel2017gans}, then by maximizing Eq.~(\ref{eq:4}), we will discover $c'$ that leads to generations of low quality. 
Recalling that our original goal is a distribution of high-quality images deviated from $p_\theta(x | c)$, we hence want to set $\varphi(\cdot):= \log p_\theta(\cdot | c)$, which, yet, is inaccessible. 
Reusing the assumption that the conditional distribution captured by a CLIP model can form a reasonable approximation to $p_\theta(x | c)$, we, in practice, set $\varphi(\cdot)$ to the aforementioned energy score $e_\phi(\cdot, c)$, which leads to the following test statistic: 
\begin{equation}
    \mathcal{D}_\text{2ST}(p_\theta(x | c') \Vert p_\theta(x | c)) := \hat{t}\Big(\{e_\phi(x_i, c)\}_{i=1}^N, \{e_\phi(\bar{x}_i, c)\}_{i=1}^N\Big).
\end{equation}
We empirically found that this t-test can lead to a superior attack success rate over other two-sample test tools, so we use the t-test as the default selection in the following.

\subsection{Attack Method}
Based on the attack objectives specified above, we define a real-world-oriented search space and employ a greedy search strategy to find adversarial input text for T2I DMs. 


\subsubsection{Perturbation Rules}
\label{sec:rules}
Following related works in natural language processing~\cite{eger-benz-2020-hero, Eger-2019-viper,le2022perturbations,DBLP:conf/emnlp/ChenGCQH0S22,chen2023adversarial}
, we include the following three kinds of perturbations into the search space of our attack algorithm: (1) \textbf{Typo}~\cite{li2018textbugger, eger-benz-2020-hero}, which comprises seven fundamental operations for introducing typos into the text, including randomly deleting, inserting, replacing, swapping, adding space, transforming case, and repeating a single character; (2) \textbf{Glyph}~\cite{li2018textbugger, Eger-2019-viper}, which involves replacing characters with visually similar ones; (3) \textbf{Phonetic}~\cite{le2022perturbations}, which involves replacing characters in a way that makes the whole word sound similar to the original one. 
We present examples of these three perturbation rules in 
Table~\ref{tab:ules}%

\begin{table}[H]
\renewcommand{\arraystretch}{1.2}
\centering
\begin{adjustbox}{max width=\textwidth}
\begin{tabular}{ccc}
\hline
\textbf{Rule} & \textbf{Ori. Sentence}     & \textbf{Adv. Sentence}      \\ \hline
Typo          & A red ball on green grass under a blue sky. & A \color[HTML]{FE0000}rde \color[HTML]{000000} ball on green grass under a blue \color[HTML]{FE0000}skky\color[HTML]{000000}.  \\ \hline
Glyph         & A red ball on green grass under a blue sky. & A \color[HTML]{FE0000}rêd \color[HTML]{000000} ball \color[HTML]{FE0000}0n \color[HTML]{000000} green grass under a blue sky.  \\ \hline
Phonetic      & A red ball on green grass under a blue sky. & A \color[HTML]{FE0000}read \color[HTML]{000000} 
 ball on green grass under a blue \color[HTML]{FE0000}SKY\color[HTML]{000000}. \\ \hline
\end{tabular}

\end{adjustbox}
\caption{Examples of our perturbation rules.}
\label{tab:ules}
\vspace{-10pt}
\end{table}

\subsubsection{Greedy Search}
\label{sec:search}
Given the efficiency and effectiveness of greedy algorithms in previous black-box text attack problems~\cite{feng2018pathologies,pruthi2019combating}, we also employ a greedy algorithm here and organize it as the following steps. 

\textbf{Step 1: word importance ranking.}
Given a sentence of $n$ words $c = \{w_1, w_2, ..., w_n\}$, it is usually the case that only some keywords act as the influential factors for controlling DMs. 
Therefore, we aim to first identify such words and then perform attack. 
The identification of word importance is trivial in a white-box scenario, e.g., by inspecting model gradients~\cite{behjati2019universal}, but is challenging in the considered black-box setting. 
To address this, we directly measure the marginal influence of the word $w_i$ on the generation distribution via $I_{w_{i}} := \mathcal{D}(p_\theta(x | c\backslash{w_i}) \Vert p_\theta(x | c))$ where
$c\backslash{w_i} = \{w_1,..., w_{i-1}, w_{i+1},... w_n\}$ denotes the sentence without the word $w_{i}$ and $\mathcal{D}$ refers to the divergence measure defined earlier. 
With this, we can compute the influence score $I_{w_{i}}$ for each word $w_i$ in the sentence $c$, and then obtain a ranking over the words according to their importance. 



\textbf{Step2: word perturbation.} 
We then attempt to perturb the detected important words to find the adversarial example $c'$. 
Concretely, for the most important word $w_i \in c$, we randomly select one character in it and then randomly apply one of the meta-operations in the perturbation rule of concern, e.g., character swapping and deleting, to obtain a perturbed word as well as a perturbed sentence. 
Repeating this five times results in 5 perturbed sentences $\{c'_1,c'_2,...c'_5\}$.
We select the sentence leading to the highest generation divergence from the original sentence, i.e., $\mathcal{D}(p_\theta(x | c'_i) \Vert p_\theta(x | c)), \forall i\in \{1,\dots,5\}$ as the eventual adversarial sentence $c'$. 
If the attack has not reached the termination condition, the next word in the importance ranking will be selected for perturbation.

\section{Diagnostic Experiments}
\label{sec:dia}





In this section, we provide diagnostic experiments consisting of two aspects: (1) assessing the four proposed attack objectives under varying perturbation rates; (2) analyzing which part of the DM is significantly misled. 
These analyses not only validate the efficacy of our method, but also deepen our understanding of adversarial attacks on T2I DMs, and offer insightful perspectives for future works.

\textbf{Dataset.} We consider two types of textual data for prompting the generation of T2I DMs: (1) 50 ChatGPT generated (ChatGPT-GP) prompts by querying: ``generate 50 basic prompts used for image synthesis.'' and (2) 50 image captions from SBU Corpus~\cite{NIPS2011_5dd9db5e}. Such a dataset facilitates a thorough investigation of the efficacy and applicability of our method in practical image-text generation tasks.

\textbf{Victim Models.}  We choose Stable Diffusion~\cite{DBLP:conf/cvpr/RombachBLEO22} as the victim model due to its widespread usage, availability as an open-source model, and strong generation capability. 
Stable Diffusion utilizes a denoising mechanism that operates in the latent space of images and incorporates cross-attention to leverage guidance information. 
Text inputs are first processed by CLIP's text encoder to generate text embeddings, which are subsequently fed into the cross-attention layers to aid in image generation.

\textbf{Evaluation Metrics.} We use the CLIP Score~\cite{hessel2021clipscore}, esentially the aforementioned $h_{\phi}(x)^\top g_{\phi}(c)$, to measure the semantic similarity between the original text $c$ and the generated images $\{x_1,\dots,x_N\}$ based on the adversarial text $c'$. 
Specifically, we define the metric $\text{$S_{I2T}$}= \frac{1}{N} \sum_{i=1}^N \max(0, 100\cdot g_{\phi}(c)^\top h_{\phi}(x'))$ over the generated images, and we hypothesize that a higher $\text{$S_{I2T}$}$ indicates a less adversarial text $c'$. 
Typically, $N$ is set to $15$ to balance efficiency and fidelity. 
We can also calculate the similarity between the original text $c$ and the adversarial text $c'$ with $\text{$S_{T2T}$}= \max(0, 100\cdot g_{\phi}(c)^\top g_{\phi}(c'))$. 
Though these two metrics use the same notations as our attack objectives, we actually use various pre-trained CLIP to instantiate them to avoid over-fitting. 
In particular, we employ the CLIP with VIT-L-patch14 backbone for attack while using VIT-L-patch14-336 for evaluation.



\subsection{Attack with Different Objectives}
\begin{figure}
  \centering
  \begin{subfigure}[b]{0.327\textwidth}
    \includegraphics[width=\textwidth]{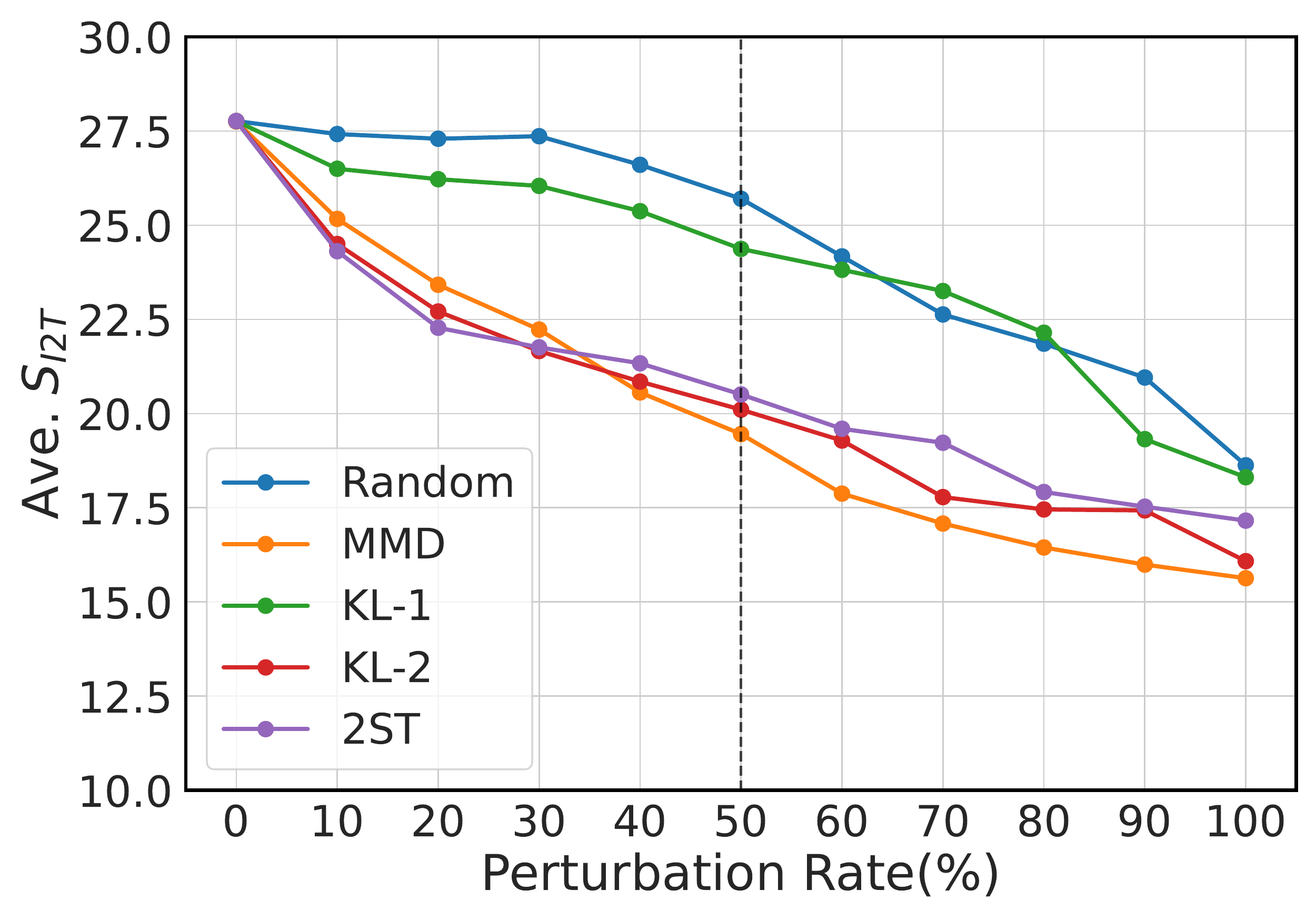}
    \caption{Typo attack on ChatGPT-GP}
  \end{subfigure}
  \begin{subfigure}[b]{0.327\textwidth}
    \includegraphics[width=\textwidth]{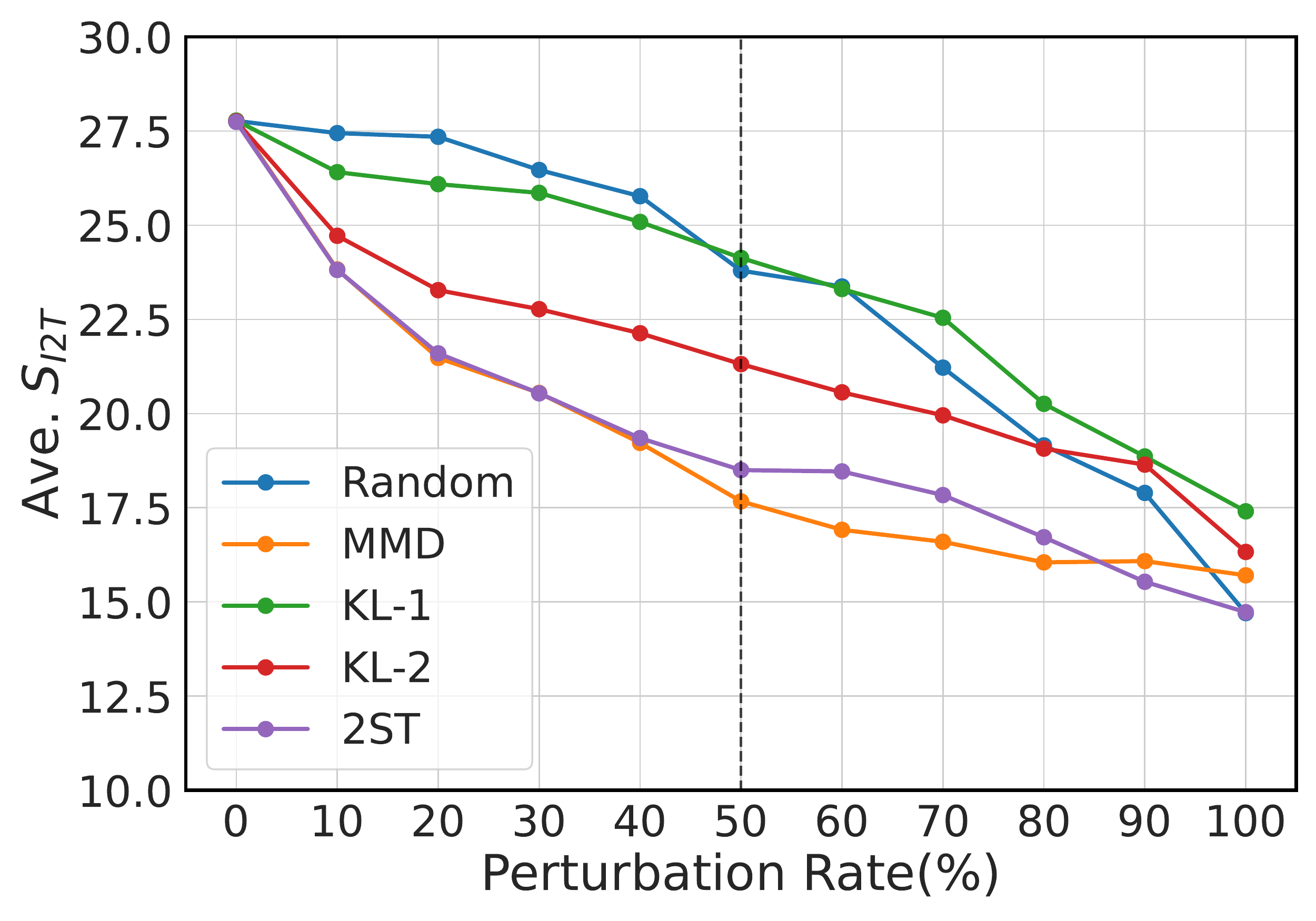}
    \caption{Glyph attack on ChatGPT-GP}
  \end{subfigure}
  \begin{subfigure}[b]{0.327\textwidth}
    \includegraphics[width=\textwidth]{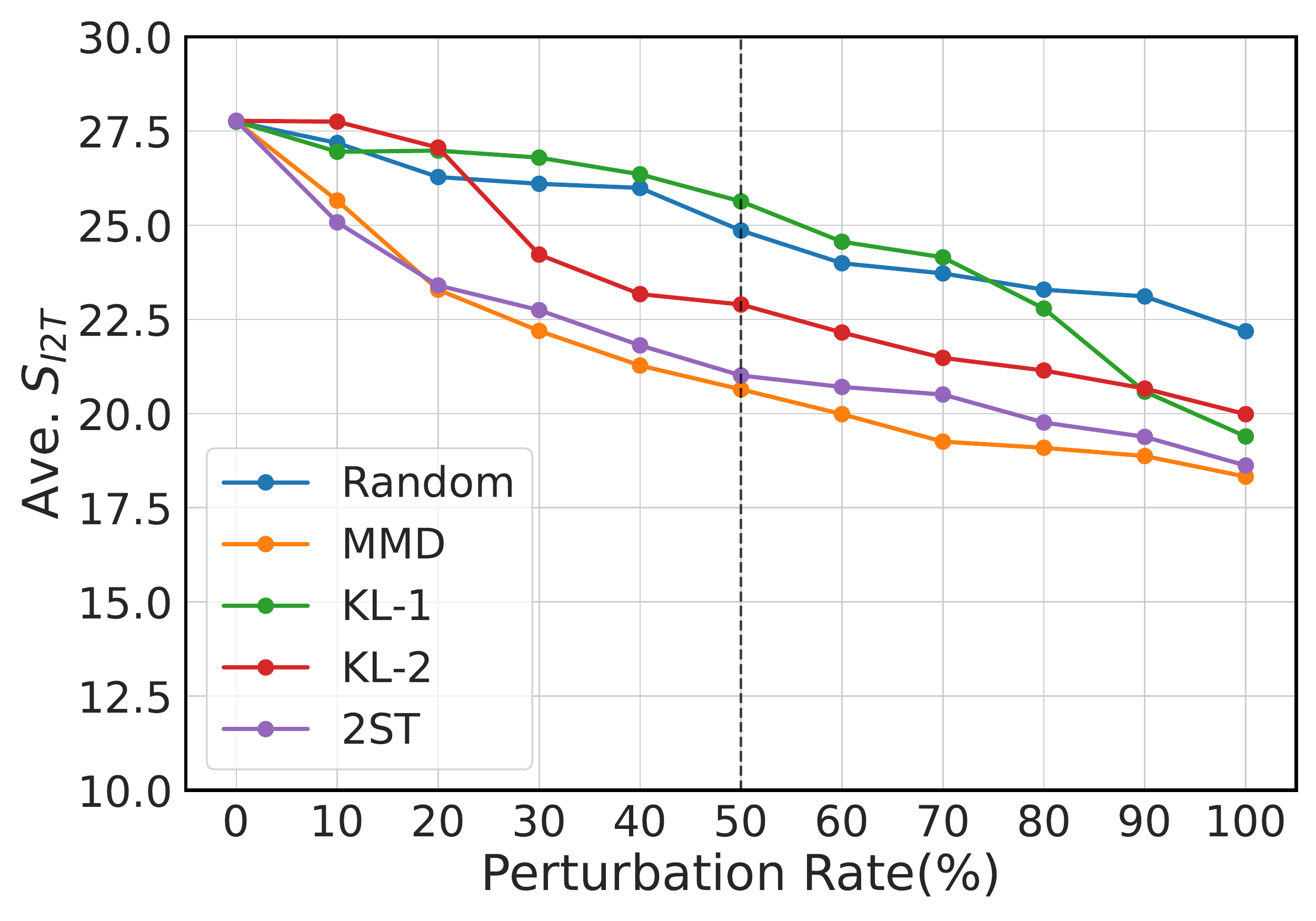}
    \caption{Phonetic attack on ChatGPT-GP}
  \end{subfigure} \
  \begin{subfigure}[b]{0.327\textwidth}
    \includegraphics[width=\textwidth]{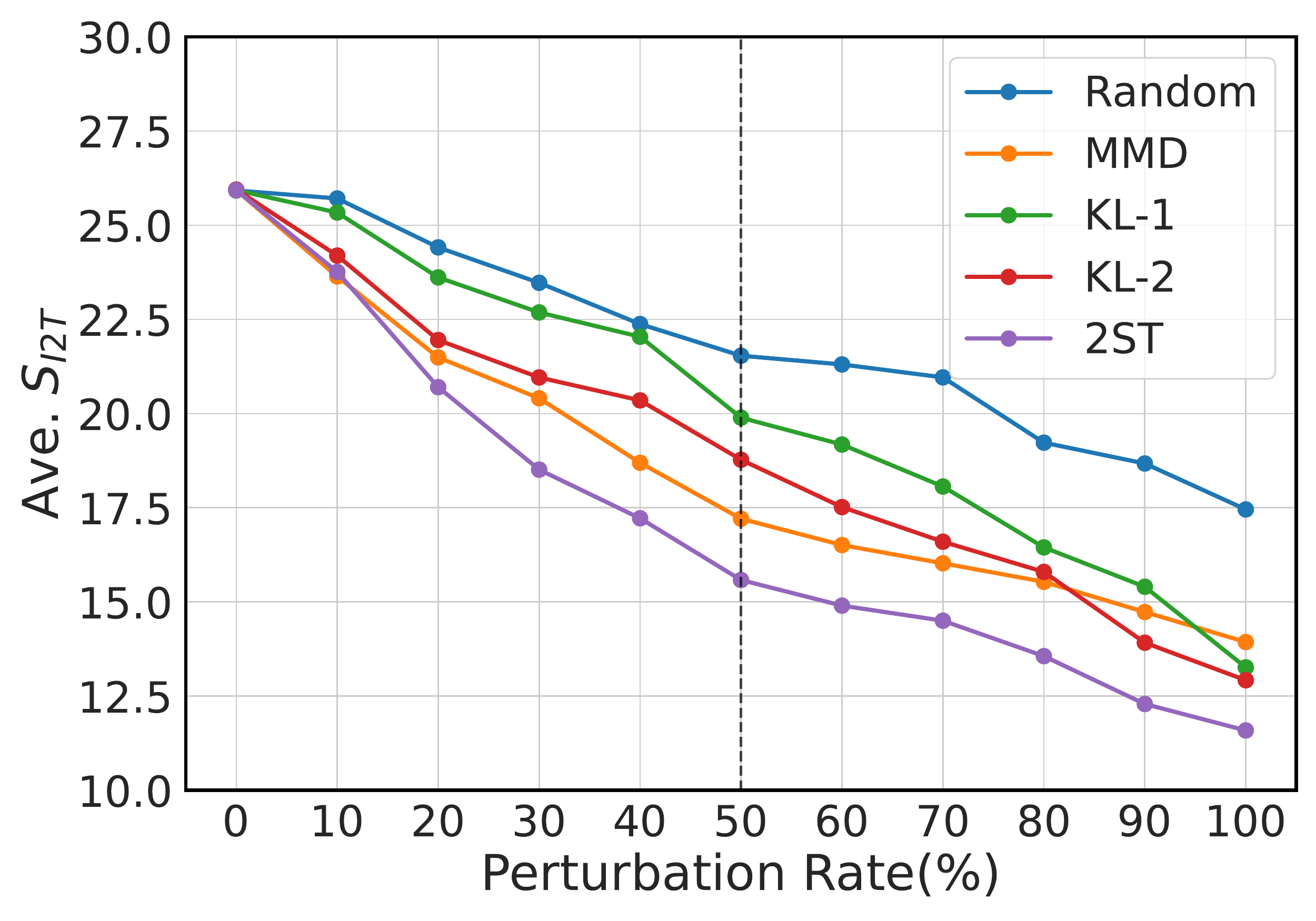}
    \caption{Typo attack on SBU Corpus}
  \end{subfigure}
  \begin{subfigure}[b]{0.327\textwidth}
    \includegraphics[width=\textwidth]{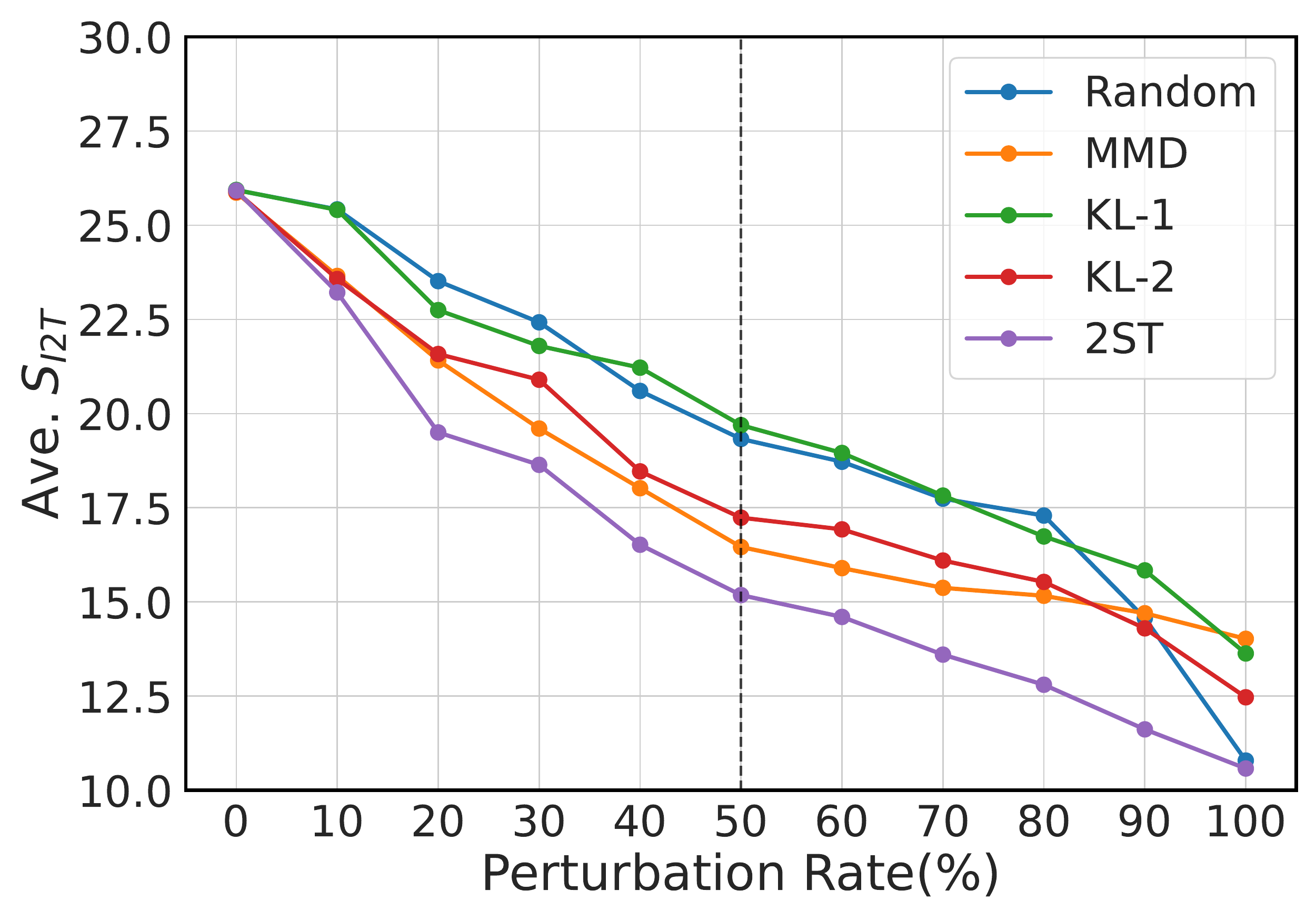}
    \caption{Glyph attack on SBU Corpus}
  \end{subfigure}
  \begin{subfigure}[b]{0.327\textwidth}
    \includegraphics[width=\textwidth]{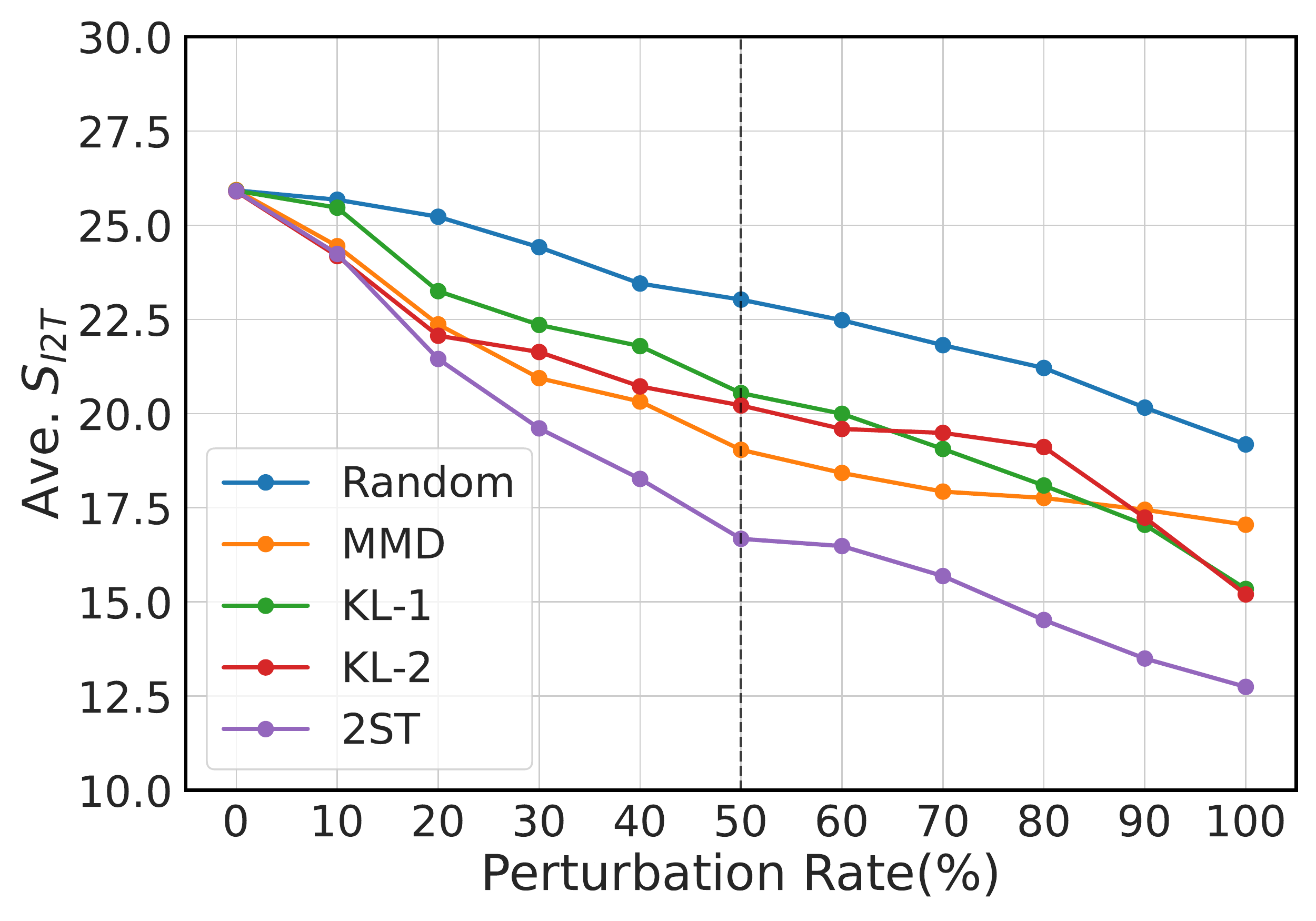}
    \caption{Phonetic attack on SBU Corpus}
    
  \end{subfigure}
  
  \caption{CLIP Score at different perturbation rates on ChatGPT-GP and SBU Corpus.}
  \label{fig:clipscore_per_chatgpt}
  \vspace{-5pt}
\end{figure}

We first conduct a diagnostic experiment on the effects of the four proposed attack objectives under various perturbation rules. 
Define the perturbation rate as the ratio between the number of perturbed words and the total words in a sentence. 
We vary it from 0\% to 100\% with an interval of 10\% and calculate the average values of $S_{I2T}$  and $S_{T2T}$ on ChatGPT-GP and SBU Corpus.
We report the results in Figure \ref{fig:clipscore_per_chatgpt}. 
Note that we also include a random baseline in comparison. 


On {ChatGPT-GP}, 
all methods exhibit a declining trend in $S_{I2T}$ as the perturbation rate increases. 
Considering high perturbation rates rarely exist in practice, we primarily focus on situations where the perturbation rate is less than 50\%. Within this range, we observe that the curves corresponding to MMD, KL-2, and 2ST display a rapid decrease across all three perturbation rules, more than $2\times$ faster than random and KL-1 when using typo and glyph rules. 
It is also noteworthy that MMD and 2ST perform similarly and yield the best overall results.

On {SBU Corpus}, 
it is evident that 2ST is more effective than MMD. 
Additionally, even with a perturbation rate of 100\%, the random method fails to achieve a similar $S_{I2T}$ score compared to other methods. 
This observation suggests the effectiveness of our 2-step attack algorithm. 
Additionally, glyph-based perturbations lead to the most rapid decrease in performance, followed by typo perturbations, and phonetic perturbations lead to the slowest drop. 
This disparity may be attributed to glyph perturbations completely disrupting the original word embedding. 

\subsection{Which Part of the DM is Significantly Misled?}
Previous studies suggest that attacking only the CLIP encoder is sufficient for misleading diffusion models~\cite{zhuang2023pilot}. 
However, our method is designed to attack the entire generation process instead of rarely the CLIP encoder. 
For empirical evaluation, we conduct a set of experiments in this section.

We include two additional attack methods: attacking only the CLIP encoder and attacking only the diffusion process. 
Regarding this first one, we focus solely on maximizing the dissimilarity between the original text and the adversarial one. To achieve this, we employ $S_{T2T}$ as the optimization objective, i.e., $\mathcal{D}_\text{CLIP} =\text{$S_{T2T}$}= \max(0, 100\cdot g_{\phi}(c)^\top g_{\phi}(c')).$
As for the second one, we modify Eq.~\eqref{eq:kl1} and devise a new attack objective as follows ($\alpha$ and $\beta$ denote two trade-off coefficients):   
\begin{equation}
\begin{aligned}
    \mathcal{D}_\text{DP} = \Big[\alpha g_{\phi}(c') - \beta \frac{1}{N} \sum_{i=1}^N h_{\phi}(x_i)\Big]^\top   g_{\phi}(c).
\end{aligned}
\end{equation}
While maximizing the distance between the original text and the adversarial images, we also aim to ensure that the representations of the adversarial text and the original text are as similar as possible. 
This confines that even though the entire DM is under attack, the CLIP encoder remains safe.  

\begin{figure}
    \centering
  \begin{subfigure}[b]{0.328\textwidth}
    \includegraphics[width=\textwidth]{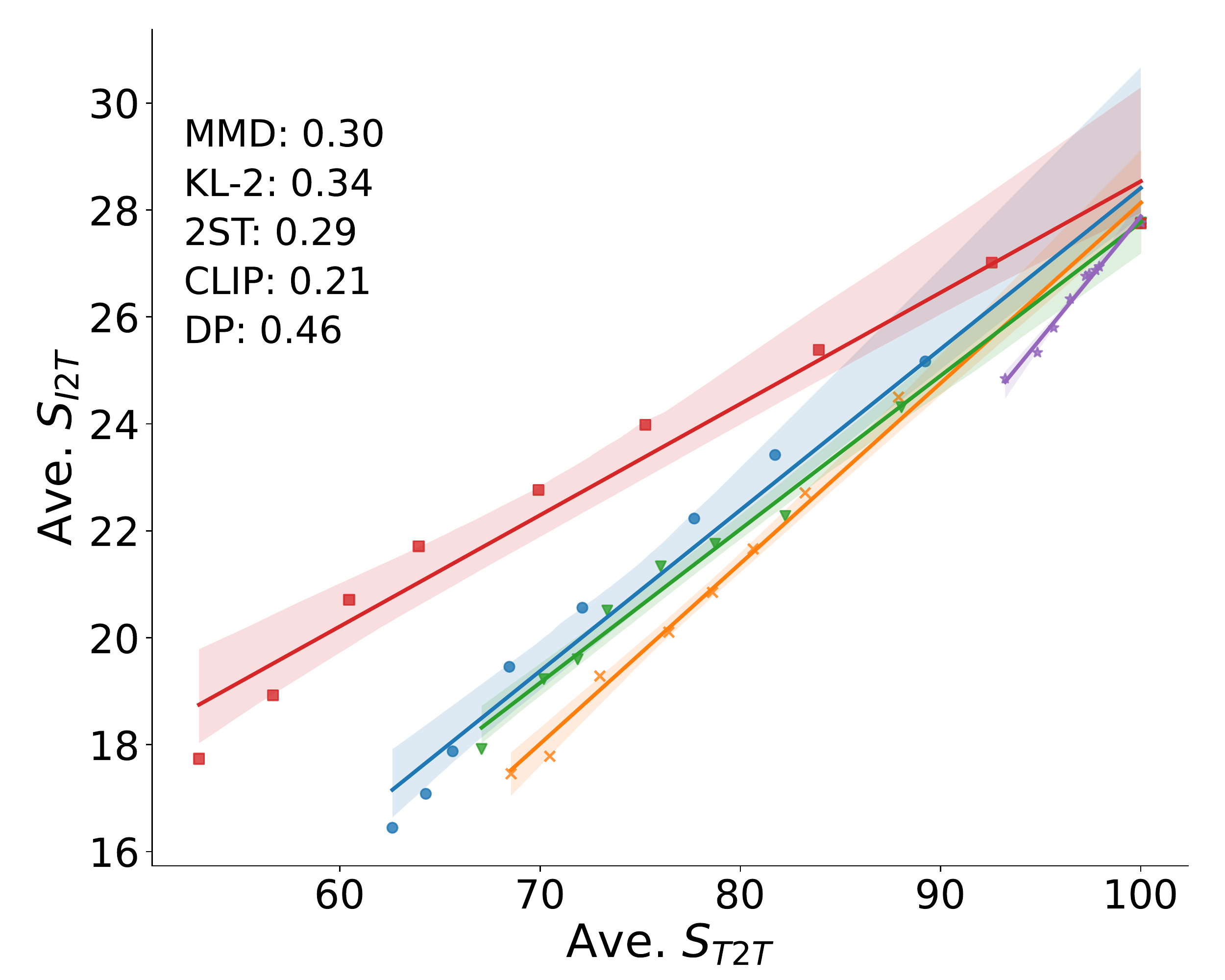}
    \caption{Typo attack on ChatGPT-GP}
  \end{subfigure}
  \begin{subfigure}[b]{0.328\textwidth}
    \includegraphics[width=\textwidth]{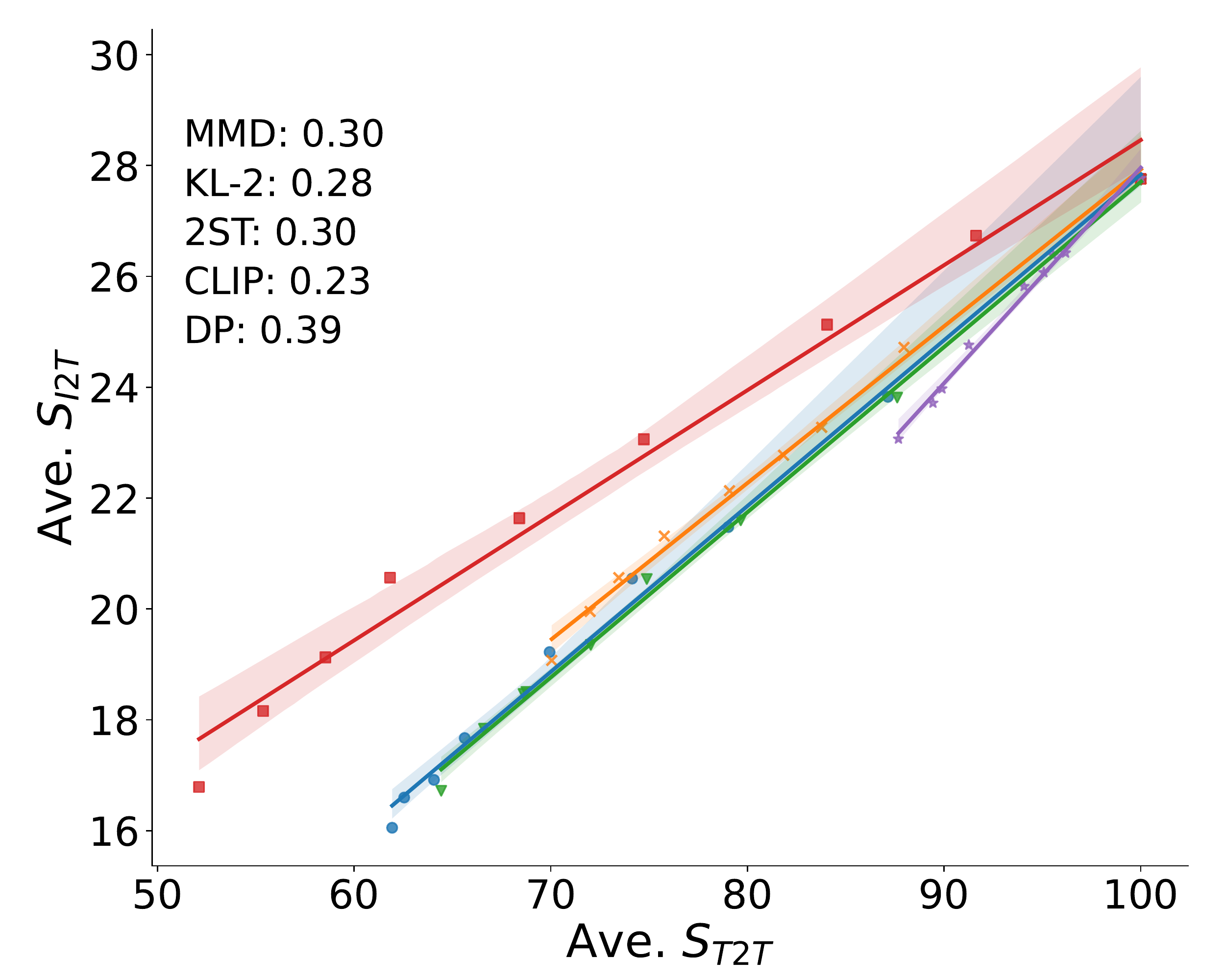}
    \caption{Glyph attack on ChatGPT-GP}
  \end{subfigure}
  \begin{subfigure}[b]{0.328\textwidth}
    \includegraphics[width=\textwidth]{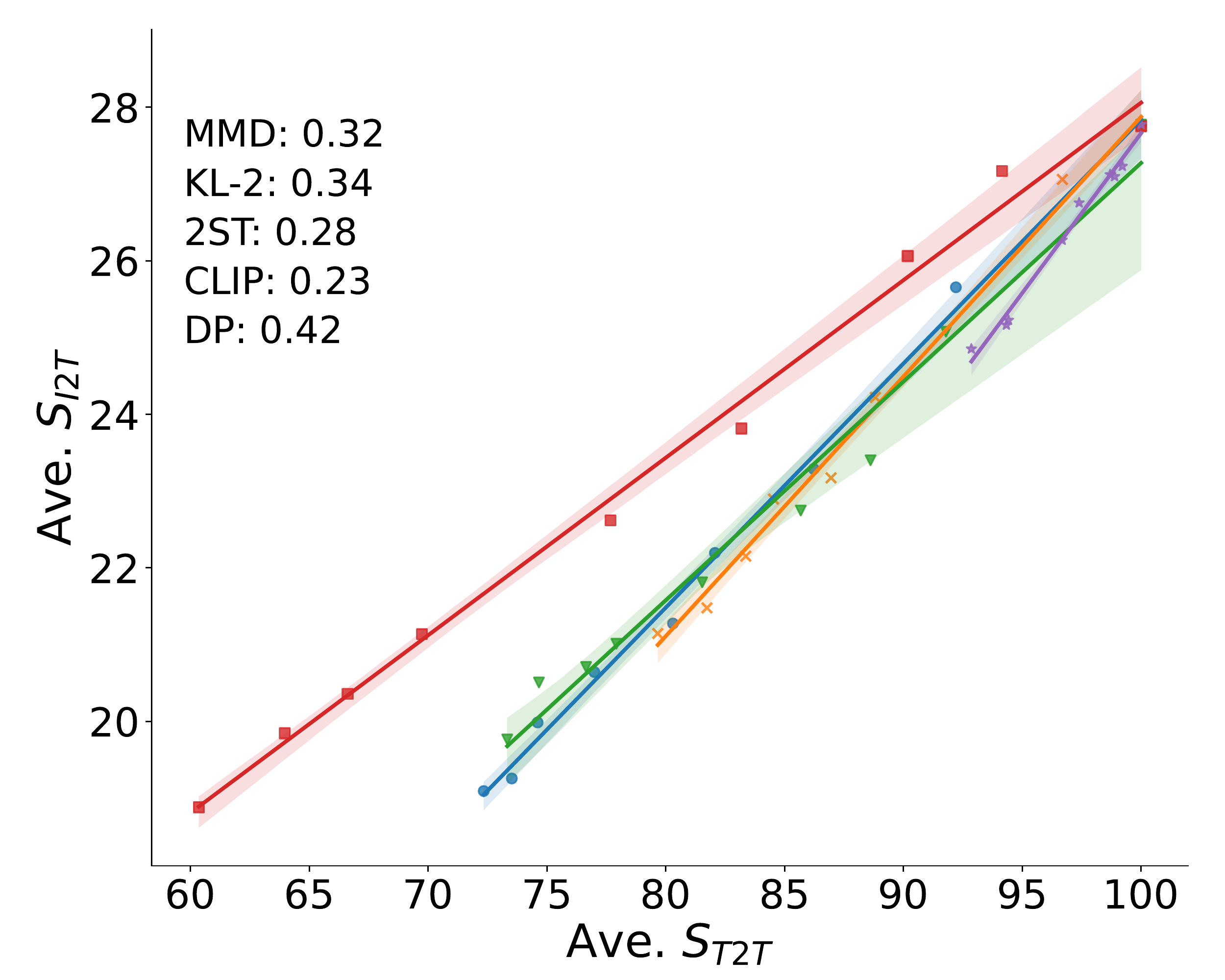}
    \caption{Phonetic attack on ChatGPT-GP}
  \end{subfigure}
    \begin{subfigure}[b]{0.328\textwidth}
    \includegraphics[width=\textwidth]{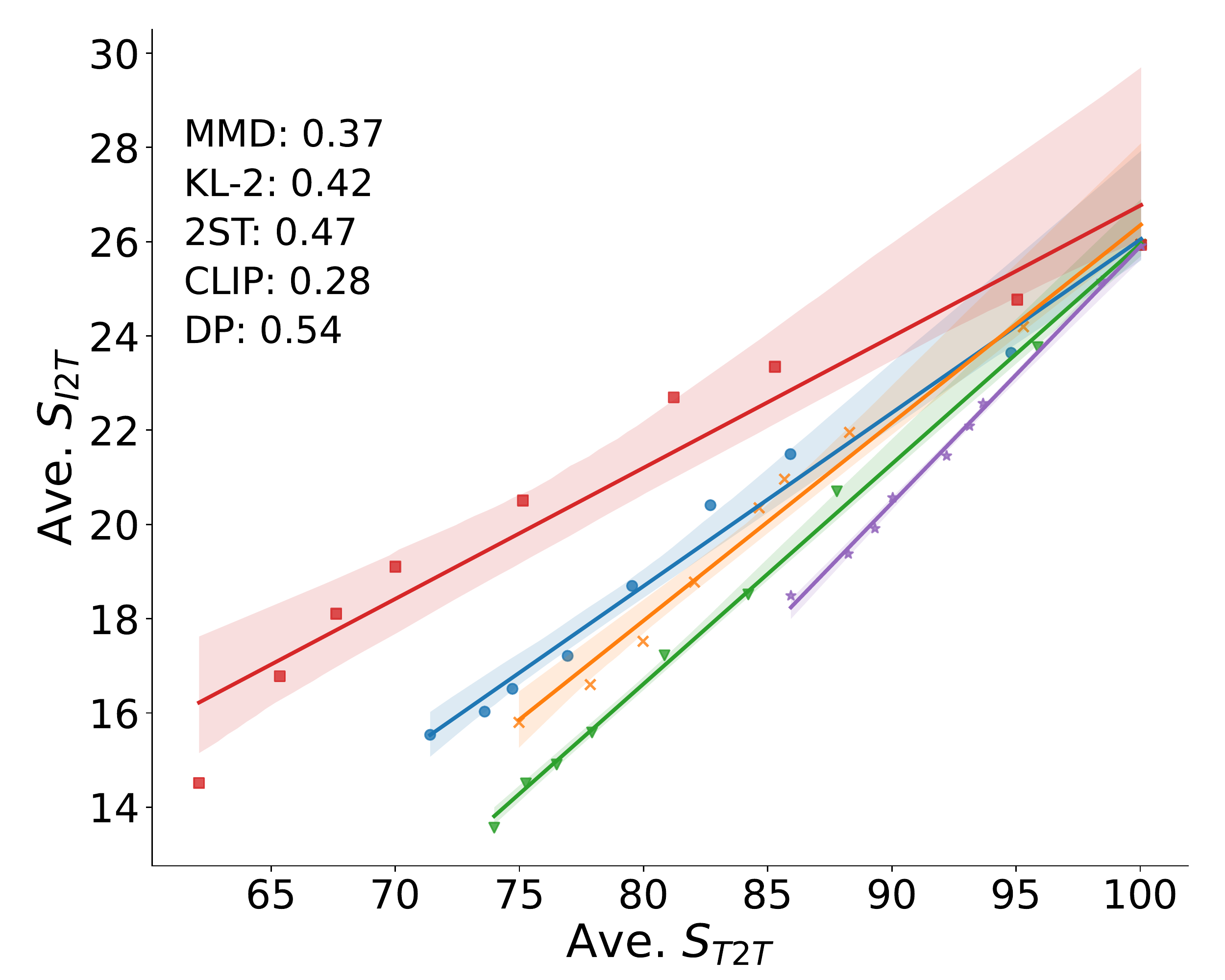}
    \caption{Typo attack on SBU Corpus}
  \end{subfigure}
  \begin{subfigure}[b]{0.328\textwidth}
    \includegraphics[width=\textwidth]{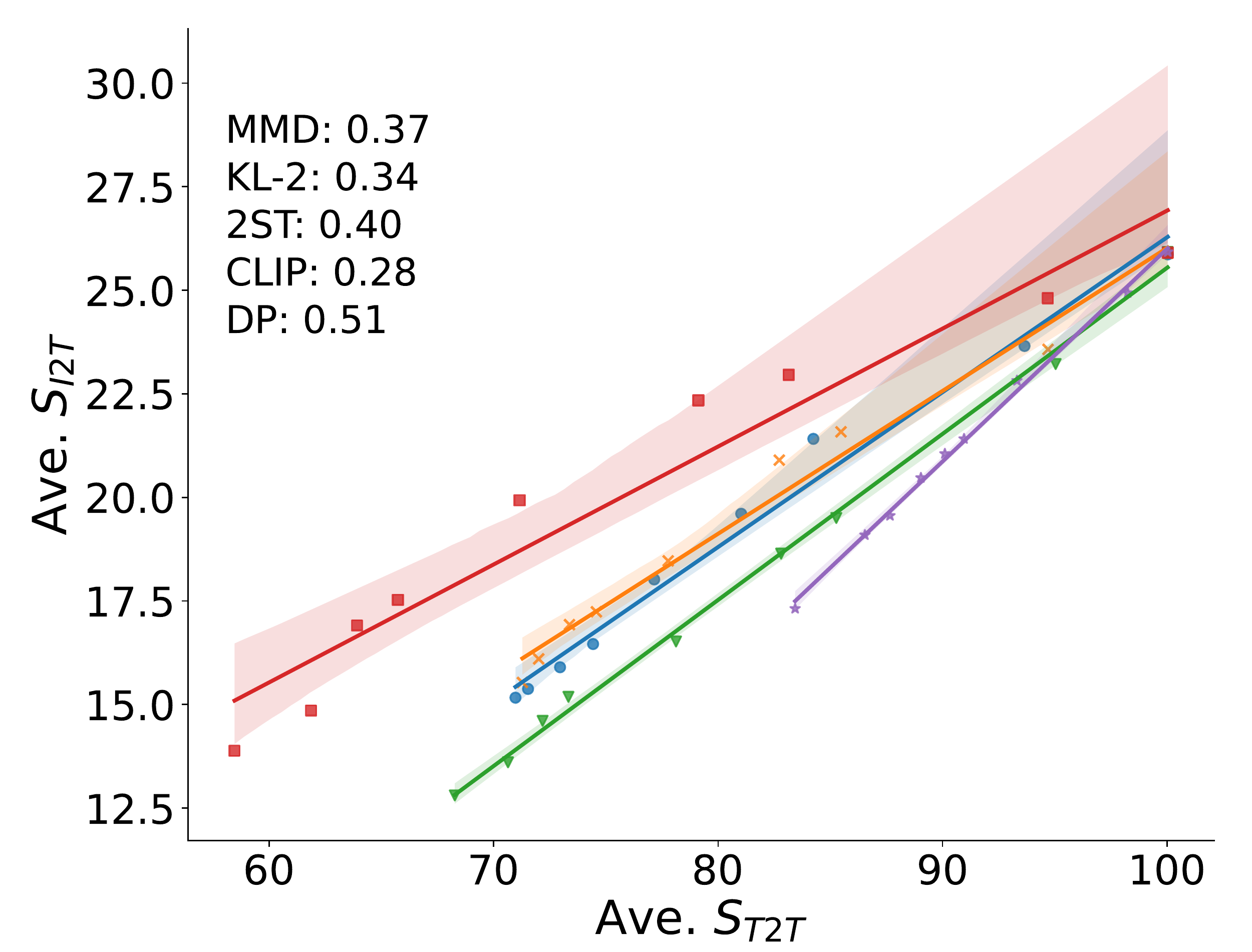}
    \caption{Glyph attack on SBU Corpus}
  \end{subfigure}
  \begin{subfigure}[b]{0.328\textwidth}
    \includegraphics[width=\textwidth]{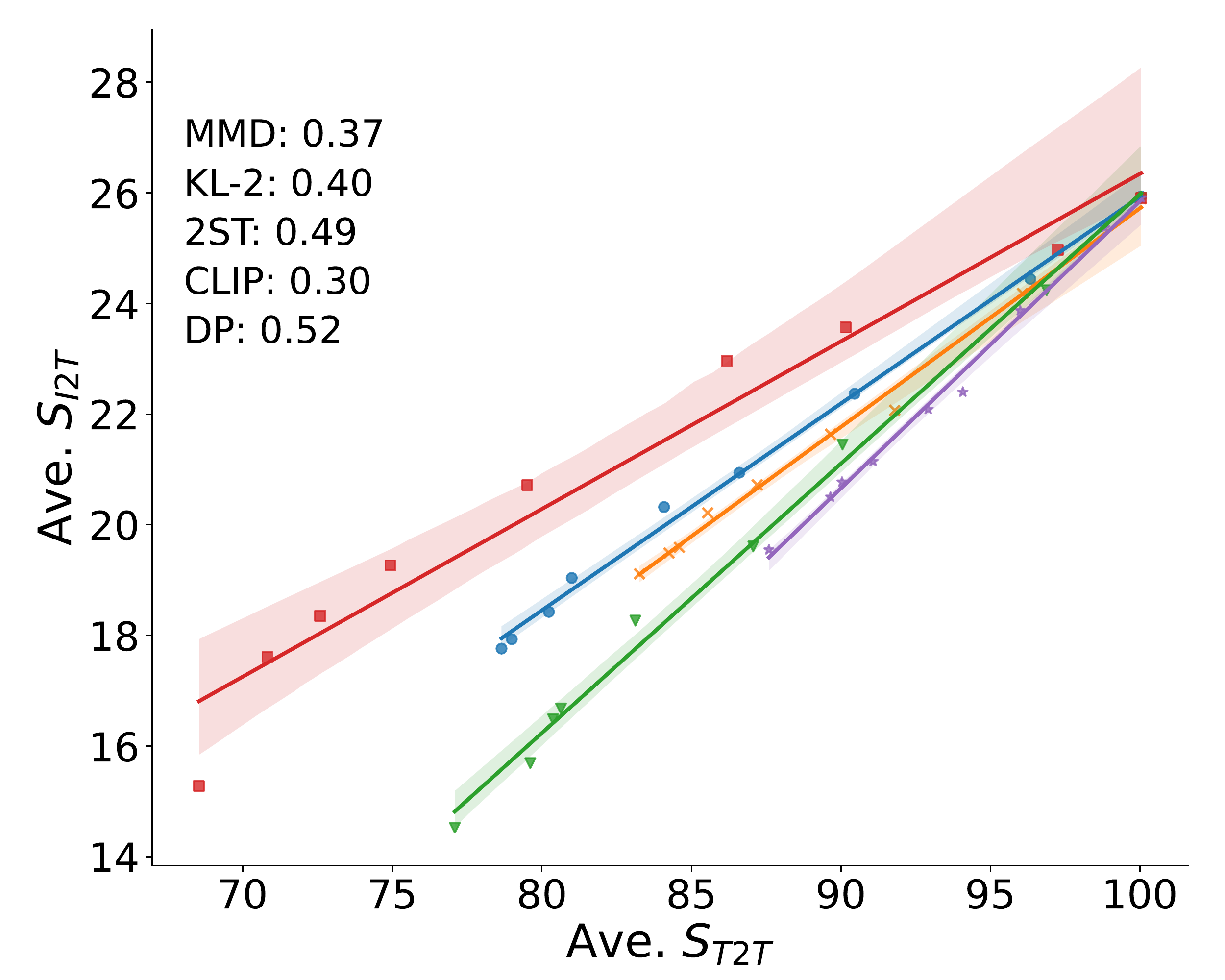}
    \caption{Phonetic attack on SBU Corpus}
  \end{subfigure}
\includegraphics[width=0.8\linewidth]{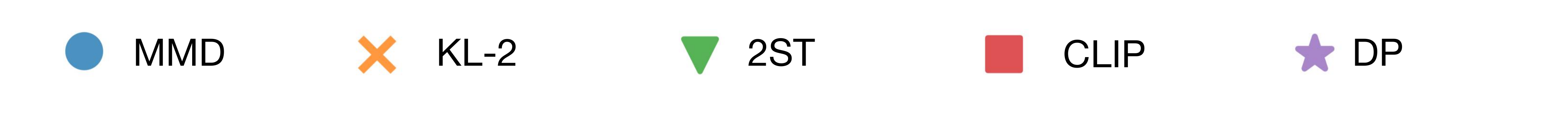}
\vspace{-5pt}
  \caption{Corelation between \text{$S_{I2T}$} and \text{$S_{T2T}$} on ChatGPT-GP and SBU Corpus. The numbers in the upper-left corner represent the slopes of the plotted lines.}
  \label{fig:clip_only_chatgpt}
  \vspace{-10pt}
\end{figure}

Given the poor performance of the random and KL-1 methods, we exclude them from this study. 
Considering that high perturbation rates are almost impossible in the real world, we experiment with perturbation rates only from 0\% to 80\%. 
We compute the average $S_{I2T}$ and $S_{T2T}$ across all texts at every perturbation rate, and plot their correlations in Figure~\ref{fig:clip_only_chatgpt}. 

As shown, exclusively targeting the CLIP encoder during the attack process yields the maximum slope of the regression line, while solely attacking the diffusion process leads to the minimum slope. 
For instance, in the typo attack on ChatGPT-GP, the attack method solely attacking the CLIP encoder exhibits the lowest slope of 0.21, whereas the attack method exclusively targeting the diffusion process shows the highest slope of 0.46. 
Attack methods that simultaneously target both processes display slopes between these extremes. 
These clearly support that our attack objectives simultaneously attack the CLIP encoder and the diffusion process. 
Furthermore, through the slope information, we can conclude that directly attacking the diffusion process yields a more significant decrease in image-text similarity at a given textual semantic divergence. 
Across all datasets and perturbation spaces, the slope disparity between direct attacks on the diffusion process and direct attacks on the CLIP encoder is mostly above 0.1, 
and the maximum slope disparity reaches even 0.15. 
\subsection{Compare with Non-distribution Attack Objective}
\begin{figure}[H]
  \centering
  \begin{subfigure}[b]{0.38\textwidth}
    \includegraphics[width=\textwidth]{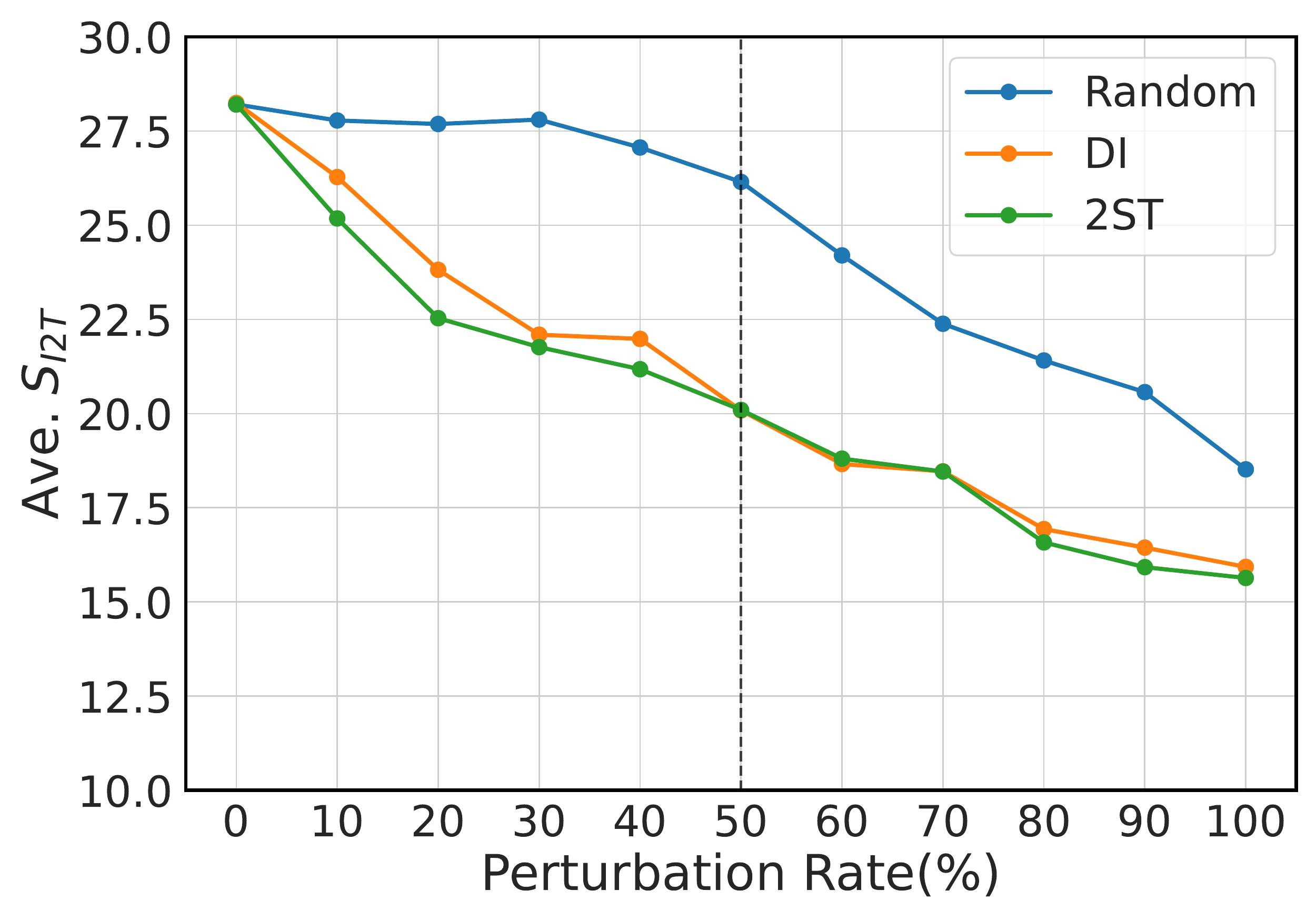}
    \caption{ChatGPT-GP}
  \end{subfigure}
  \begin{subfigure}[b]{0.38\textwidth}
    \includegraphics[width=\textwidth]{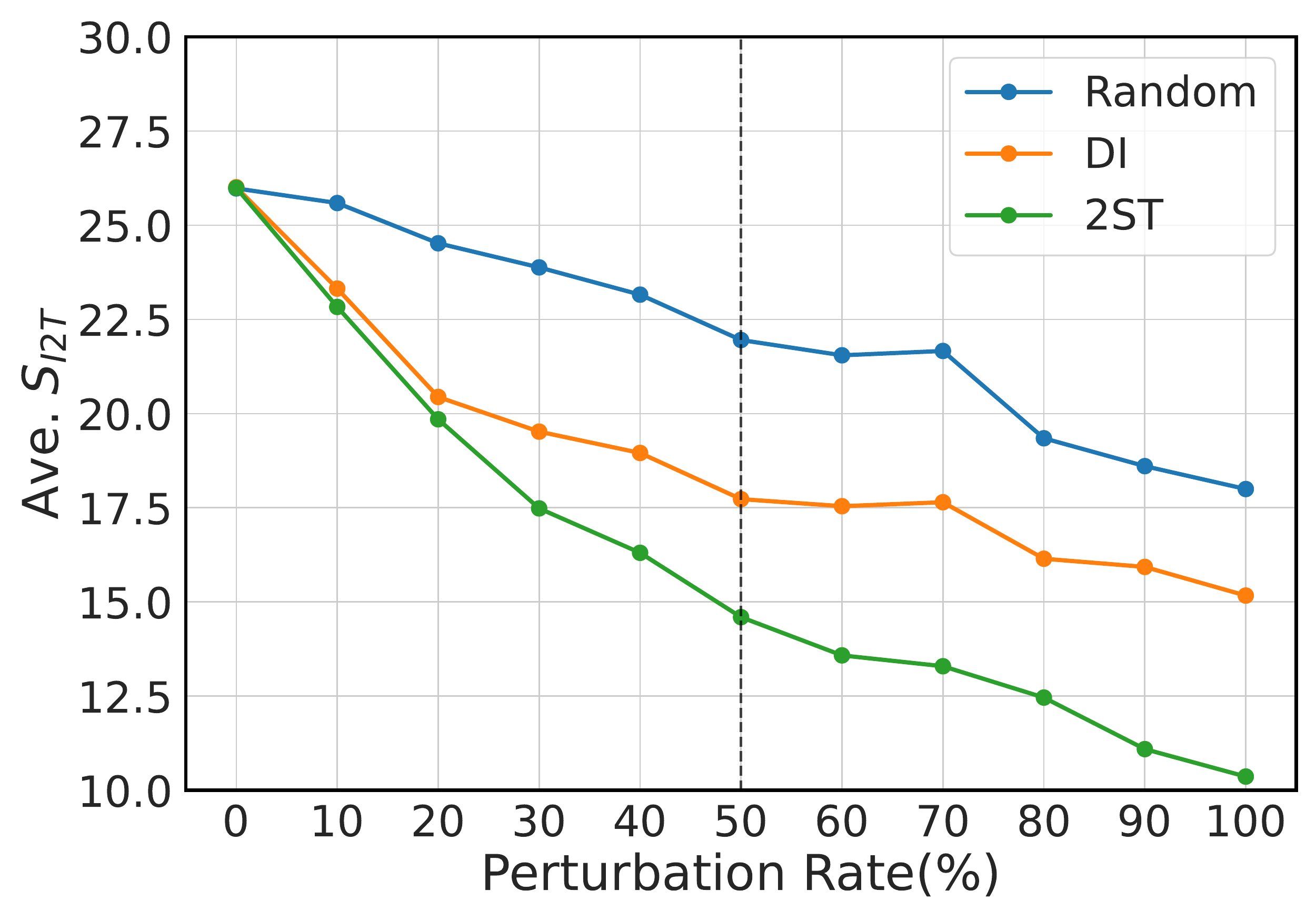}
    \caption{SBU Corpus}
  \end{subfigure}
    
  
  \caption{CLIP Score at different perturbation rates on ChatGPT-GP and SBU Corpus with typo rule.}
  \label{ast}
  \vspace{-10pt}
\end{figure}

We conducted a comparison experiment between our distribution-based optimization objective, referred to as \textbf{2ST}, and a non-distribution method that solely relies on the $S_{T2I}$ of the prompt combined with a single definite image (DI). Following Section 4, we randomly sampled 20 texts from ChatGPT-GP and SBU Corpus separately, then applied typo rule to perturb sampled texts with different perturbation rates. The results, depicted in Figure~\ref{ast}, clearly demonstrate the superior effectiveness of the distribution-based approach.

\section{Real-world Attack Experiment}
\subsection{Stable Diffusion}
Based on the preceding analysis, we identify that 2ST and MMD are two good attack objectives for T2I DMs. 
In this section, we will carry out attacks in real-world scenarios, where termination conditions are incorporated to balance the perturbation level and effectiveness. 

\textbf{Datasets.} 
To provide a more comprehensive evaluation of our attack method in realistic scenarios, we incorporate two additional datasets. 
The first one, DiffusionDB~\cite{wangDiffusionDBLargescalePrompt2022}, is a large-scale dataset of 14 million T2I prompts. 
The second one, LAION-COCO~\cite{schuhmann2021laion}, includes captions for 600 million images from the English subset of LAION-5B~\cite{schuhmannlaion}. 
The captions are generated using an ensemble of BLIP L/14~\cite{li2022blip} and two CLIP~\cite{radford2021learning} variants. 
To conjoin diversity and efficiency, we randomly selecte 100 examples from each of the aforementioned datasets. 
Additionally, we also increase the size of  ChatGPT-GP and SBU to 100 for this experiment.

\textbf{Attack method.} 
As said, we consider attacking based on MMD and 2ST. 
A threshold on the value of $\mathcal{D}$ is set for termination. 
If it is not reached, the attack terminates at a pre-fixed number of steps.

\textbf{Evaluation metric.} We use four metrics to evaluate our method in real-world attack scenes. (1) Levenshtein distance (L-distance), which measures the minimum number of single-character edits, a powerful indicator of the number of modifications made to a text. (2) Ori.$S_{I2T}$ and Adv.$S_{I2T}$ which indicate the similarity between the original text and original images as well as that between the original text and the adversarial images respectively. The mean and variance are both reported. (3) Average query times, which represents the number of times that DM generates images with one text, and serves as a metric for evaluating the attack efficiency. (4) Human evaluation, where humans are employed to assess the consistency between the image and text. 
Let $N_1$ represent the consistency between the original text and the original image, and $N_2$ represent the consistency between the original text and the adversarial image. If ($N_2 - N_1$) > 1, the attack on that particular prompt text is deemed meaningless. Let's assume the frequency of samples where ($N_2 - N_1$) > 1 as $N_u$, and the effective total number of samples should be $N_{\text{total}} - N_u$. If $(N_1 - N_2 > 1)$, it indicates a successful attack. We use $N_c$ to represent the number of samples where the attack is successful. Thus, the final score for each evaluator is given by $N_c / (N_{\text{total}} - N_u)$. The average of three human annotators represents the overall human evaluation score (Hum.Eval).

\begin{table}[t!]
\renewcommand{\arraystretch}{1.2}
\centering
\begin{adjustbox}{max width=\textwidth}
\begin{tabular}{cccccccc}

\hline\hline
\textbf{Dataset} & \textbf{Attacker} & \textbf{Ori. $\mathbf{S_{I2T}}$} & \textbf{Adv. Len.} & \textbf{L-distance} & \textbf{Adv. $\mathbf{S_{I2T}}$} & \textbf{Ave. Query} & \textbf{Hum. Eval.} \\ \hline
                 & Typo                   &                          &                    & 2.92                & 23.21±3.08               & 19.43     &84.34\%          \\
ChatGPT-GP       & Glyph                  & 27.61±2.07               & 10.41              & 2.27                & 23.09±2.75               & 18.63    & 84.65\%          \\
                 & Phonetic               &                          &                    & 5.38                & 22.67±3.58               & 17.78    & 86.16\%           \\ \hline
                 & Typo                   &                          &                    & 2.29                & 22.70±3.31                & 17.25     &76.64\%          \\
DiffusionDB      & Glyph                  & 29.17±3.36               & 10.62              & 1.81                & 22.71±3.22               & 16.30      &76.64\%         \\
                 & Phonetic               &                          &                    & 5.04                & 22.91±3.34               & 16.27     &75.51\%          \\ \hline
                 & Typo                   &                          &                    & 2.08                & 21.73±3.62               & 14.77      &80.21\%         \\
LAION-COCO        & Glyph                  & 27.54±2.86               & 9.17               & 1.85                & 21.32±3.69               & 15.11        &81.89\%       \\
                 & Phonetic               &                          &                    & 5.04                & 21.76±3.87               & 16.15          &79.32\%     \\ \hline
                 & Typo                   &                          &                    & 2.97                & 19.65±3.53               & 21.19          &84.34\%     \\
SBU Corpus              & Glyph                  & 24.99±3.43               & 11.69              & 2.42                & 19.01±3.76                & 20.54           &85.41\%    \\
                 & Phonetic               &                          &                    & 5.85                & 18.86±3.91               & 19.92           &85.41\%    \\ \hline
\end{tabular}
\end{adjustbox}
\caption{\label{tab:t_test}Real-world attack with the \textbf{2ST} attack objective.}
\vspace{-10pt}
\end{table}

Table~\ref{tab:t_test} and 
Table~\ref{tab:mmd} 
present the results of our real attack experiments using various perturbation rules on different datasets, with 2ST and MMD distance as the attack objectives, respectively. Since the termination criteria for the two optimization algorithms differ, we cannot compare them directly. Considering that our method involves querying each word of the sentence (described in Section~\ref{sec:search}), the query times minus the sentence length, which we named \emph{true query times}, can better demonstrate the true efficiency of our approach. From this perspective, our method requires less than 10 \emph{true query times} to achieve more than 4 $S_{I2T}$ score drop across most datasets. 
The human evaluation score is no less than $75\%$. 
Simultaneously, we observe that our modifications are relatively minor. 
In the typo and glyph attacker, we require an L-distance of less than 3, while in the Phonetic attacker, the threshold remains below 6. Furthermore, ChatGPT-GP and LAION-COCO are more susceptible to our attack, possibly attributed to their clearer sentence descriptions and improved flow in the text. 
In conclusion, with minimal modifications and a limited number of queries to the model, we achieve a significant decrease in text-image similarity, substantiated by human evaluations.

\begin{table}[t!]
\renewcommand{\arraystretch}{1.2}
\centering
\begin{adjustbox}{max width=\textwidth}
\begin{tabular}{cccccccc}
\hline\hline
\textbf{Dataset} & \textbf{Attacker} & \textbf{Ori. $\mathbf{S_{I2T}}$} & \textbf{Adv. Len.} & \textbf{L-distance} & \textbf{Adv. $\mathbf{S_{I2T}}$} & \textbf{Ave. Query} & \textbf{Hum. Eval.} \\ \hline
                 & Typo              &                          &                    & 1.77                & 24.54±2.69               & 14.17               & 84.21\%           \\
ChatGPT-GP       & Glyph             & 27.61±2.07               & 10.41              & 1.15                & 24.88±2.67               & 13.08               & 84.36\%         \\
                 & Phonetic          &                          &                    & 3.81                & 26.08±2.21               & 14.58               & 80.02\%            \\ \hline
                 & Typo              &                          &                    & 1.75                & 24.94±3.82               & 13.72               & 72.77\%           \\
DiffusionDB      & Glyph             & 29.17±3.36               & 10.62              & 1.29                & 24.81±3.90               & 13.41               & 73.53\%           \\
                 & Phonetic          &                          &                    & 4.27                & 26.71±3.24               & 15.13               & 70.09\%             \\ \hline
                 & Typo              &                          &                    & 1.75                & 23.04±4.10                & 13.33               & 80.21\%           \\
LAION-COCO        & Glyph             & 27.54±2.86               & 9.17               & 1.35                & 23.72±3.91               & 12.35               & 82.04\%           \\
                 & Phonetic          &                          &                    & 3.62                & 25.06±3.09               & 13.21               & 77.37\%           \\ \hline
                 & Typo              &                          &                    & 1.91                & 21.37±3.92               & 16.36               & 82.05\%           \\
SBU Corpus              & Glyph             & 24.99±3.43               & 11.69              & 1.37                & 21.44±3.66               & 15.01               & 82.33\%           \\
                 & Phonetic          &                          &                    & 3.72                & 23.15±3.25               & 16.20                & 79.67\%         \\ \hline
\end{tabular}
\end{adjustbox}
\caption{\label{tab:mmd}Real-world attack with \textbf{MMD distance} attack objective.}
\vspace{-20pt}
\end{table}

\subsection{DALL-E 2}
\begin{figure}[H]
    \centering
        \includegraphics[page=2,width=\textwidth]{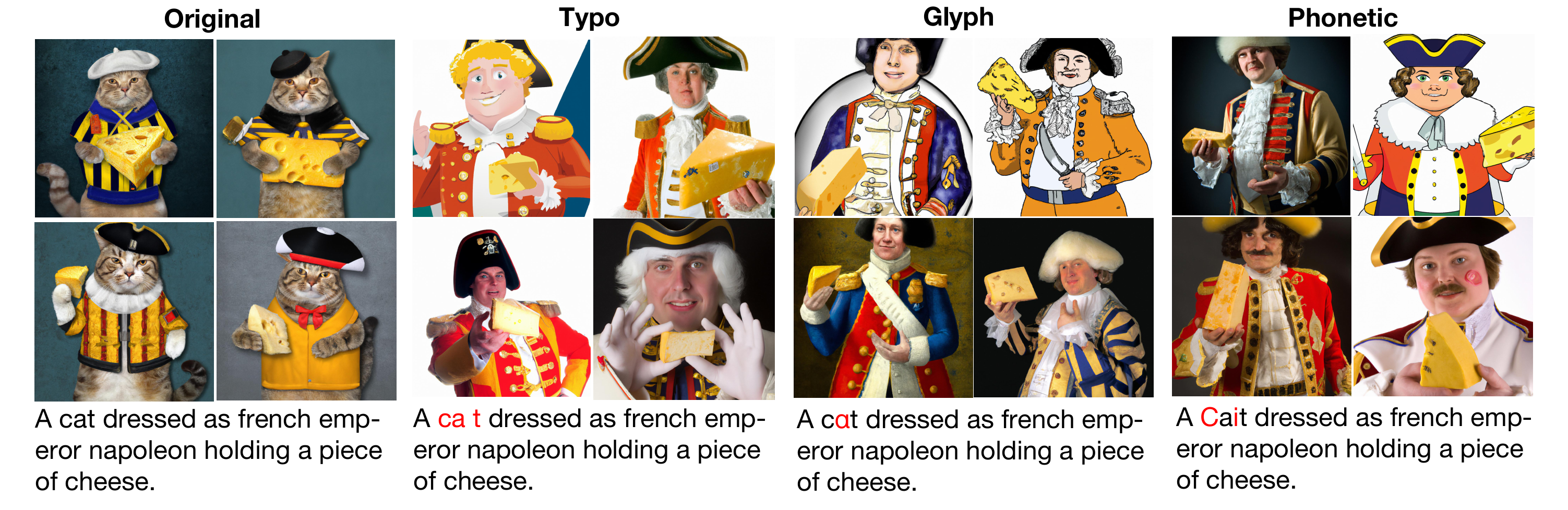}
  \caption{An illustration of adversarial attack against DALL-E 2 with \textbf{MMD} attack objective. }
  \vspace{-10pt}
  \label{dalle_example1}
\end{figure}

\begin{figure}[H]
    \centering
        \includegraphics[page=1,width=\textwidth]{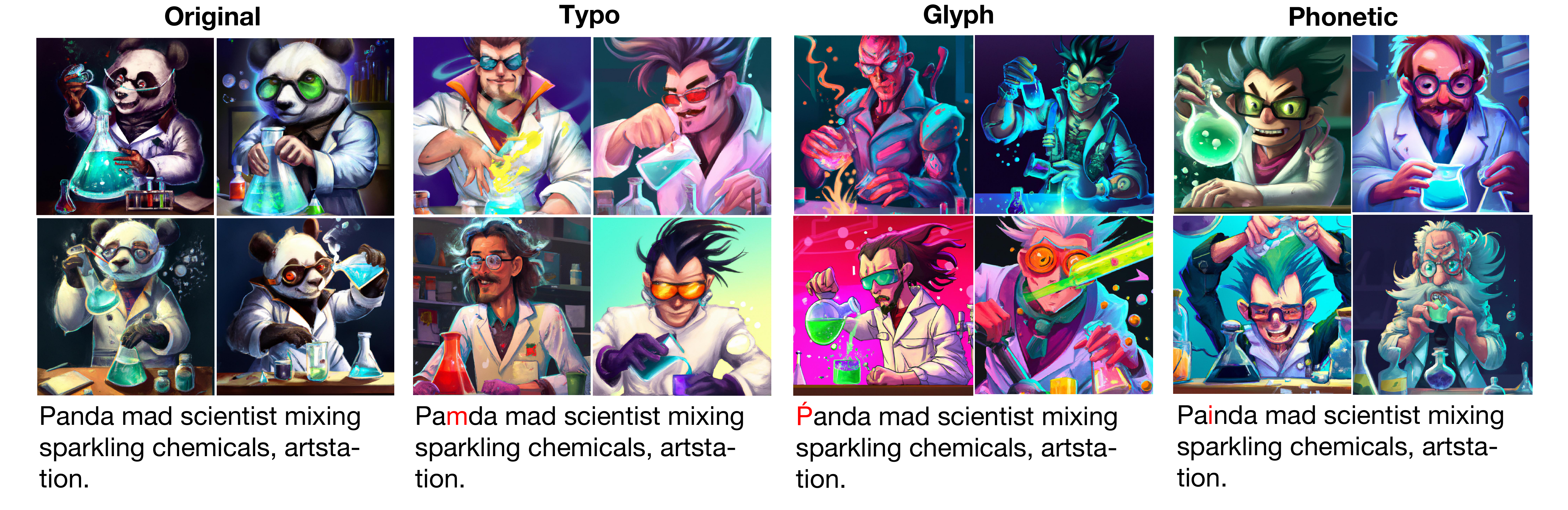}
  \caption{An illustration of adversarial attack against DALL-E 2 with \textbf{2ST} attack objective. }
  \vspace{-10pt}
  \label{dalle_example2}
\end{figure}

DALL-E 2 is a powerful image generation model that can create realistic and diverse images from textual descriptions. We conduct a case study with the same attack method used in Stable Diffusion. The results respectively obtained with the attack objective MMD and 2ST are presented in Figure~\ref{dalle_example1} and Figure~\ref{dalle_example2}.

\section{Conclusion} 
In this work, we present a comprehensive evaluation of the robustness of DMs against real-world attacks. Unlike previous studies that focused on malicious alterations to input texts, we explore an attack method based on realistic errors that humans can make to ensure semantic consistency. Our novel distribution-based attack method can effectively mislead DMs in a black-box setting without any knowledge of the original generative model. 
Importantly, we show that our method does not solely target the text encoder in DMs, it can also attack the diffusion process. 
Even with extremely low perturbation rates and query times, our method can achieve a high attack success rate. 

In appendix~\ref{sec:discuss}, we discuss the challenges encountered in attacking without algorithmic interventions, providing evidence for the effectiveness of our attack algorithm. Additionally, we explain why this study did not employ word-level attacks and highlight the impracticality of directly transferring text-based adversarial attacks to DMs.

\section{Limitation and Border Impact}
\textbf{Limitation.} In our experiments, we employ DMs as the testbed and evaluate both random attack methods and our proposed method with four optimization objectives on our custom benchmark datasets. Due to limited resources, we focus on Stable Diffusion for the complete experiment and DALL-E 2 for the case study, given that our method involves 12 combinations. Therefore, conducting more comprehensive experiments covering different model architectures and training paradigms is a direction for future research.

\textbf{Border Impact.} A potential negative societal impact of our approach is that malicious adversaries could exploit it to construct targeted attacks by modifying the loss function, leading to the generation of unhealthy or harmful images, thus causing security concerns. As more people focus on T2I DMs due to their excellent performance on image generation. In such scenarios, it becomes inevitable to address the vulnerability of DMs which can be easy attack through black-box perturbation. Our work emphasizes the importance for developers of DMs to consider potential attacks that may exist in real-world settings during the training process.

\newpage
{\small
\bibliographystyle{unsrt}
\bibliography{main}
}

\appendix
\newpage
\section{Proof of Eq.~(\ref{eq:kl})}
\begin{equation}
\begin{aligned}
\centering
    \mathcal{D}_\text{KL}(p_\theta(x | c') \Vert p_\theta(x | c)) \approx \, &   \mathcal{D}_\text{KL}(p_\theta(x | c') \Vert p_\phi(x | c))\\
    =\, & \mathbb{E}_{p_\theta(x | c')} \Big[\log \frac{p_\theta(x | c')}{p_\phi(x | c) p(c)} + \log p(c)\Big] \\
    =\, & \mathbb{E}_{p_\theta(x | c')} [-{ e_\phi(x, c)}] + \mathbb{E}_{p_\theta(x | c')} [\log {p_\theta(x | c')}]  + C \\
\end{aligned}
\end{equation}

\section{Discussion and Future Work}
\label{sec:discuss}

\begin{figure}[H]
    \centering
        \includegraphics[page=6,width=\textwidth]{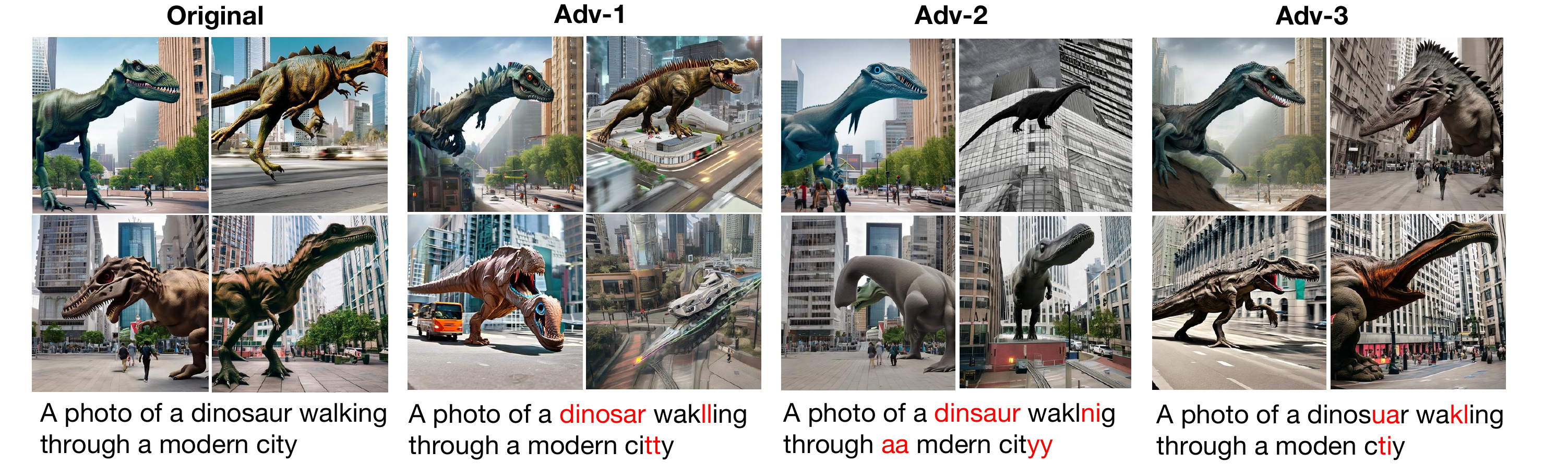}
  \caption{An illustration of human attack against Stable Diffusion. Adversarially modified content is highlighted in \color[HTML]{FE0000} red \color[HTML]{000000}.}
  \label{image_example1}
\end{figure}

\begin{figure}[H]
    \centering
        \includegraphics[width=\textwidth]{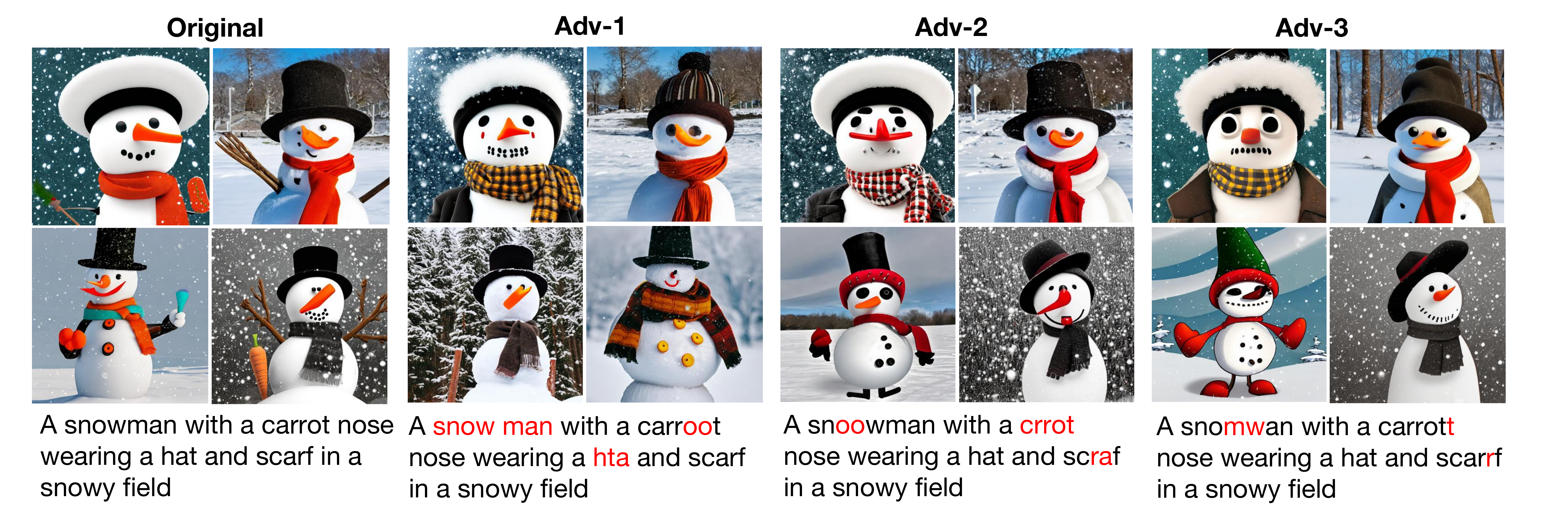}
  \caption{An illustration of human attack method against Stable Diffusion. Adversarially modified content is highlighted in \color[HTML]{FE0000} red \color[HTML]{000000}.}
  \label{image_example2}
\end{figure}

Firstly, we would like to emphasize the effectiveness of our adversarial optimization algorithm. In order to demonstrate this, we randomly selected a set of sentences and made random modifications to the most important words based on human intuition. Remarkably, we observed that a lot of sentences with these modifications did not result in DMs generating incorrect images. This further substantiates the meaningfulness of our attack algorithm. We present two illustrative cases in Figure~\ref{image_example1} and Figure~\ref{image_example2}.

Then we talk about the other level attack such as word-level attack. Due to the high sensitivity of the DM to individual words in the prompt, word-level attacks such as synonym replacement or context filling were not employed in this study. If we were to use synonym replacements and substitute words with less commonly used ones, those words themselves might have multiple meanings. In such cases, the model is likely to generate images based on alternative meanings, making the substituted words different in the context of the sentence, even though they may be synonyms in terms of individual words. Therefore, a more stringent restriction is required for word-level replacements. It is precisely because of this reason that traditional text-based attack methods are not applicable to image-text generation. For instance, in sentiment classification tasks, they only consider the overall sentiment of the entire sentence, and the ambiguity of a particular word does not significantly impact the overall result. However, this sensitivity becomes crucial in the context of T2I. Hence, further research is needed to explore word-level and sentence-level attacks on T2I generation models. We list some examples generated by some word-level adversarial attack methods  of natural language processing(NLP) with our proposed optimization objective in table~\ref{tab:word}. It is evident that significant semantic changes have occurred in the examples presented. Both word-level and sentence-level attacks still have a long way to go in T2I adversarial attack.

\begin{table}[H]
\centering
\renewcommand{\arraystretch}{1.3}
\begin{adjustbox}{max width=\textwidth}

\begin{tabular}{lll}
\hline
\textbf{Attack Method} & \textbf{Ori. Text}                                                      & \textbf{Adv. Text}                                                      \\ \hline
              & A red ball on green grass under a blue sky.                    & A red  \color[HTML]{FE0000}field \color[HTML]{000000} on green grass under a blue sky.                   \\
BERTAttack    & A white cat sleeping on a windowsill with a flower pot nearby. & A \color[HTML]{FE0000}green \color[HTML]{000000} cat sleeping on a windowsill with a flower pot nearby. \\
              & A wooden chair sitting in the sand at a beach.                 & A wooden \color[HTML]{FE0000}camera \color[HTML]{000000} sitting in the sand at a beach.                \\ \hline
              & A red ball on green grass under a blue sky.                    & A red \color[HTML]{FE0000}orchis \color[HTML]{000000} on green grass under a blue sky                   \\
PWWS          & A white cat sleeping on a windowsill with a flower pot nearby. & A white \color[HTML]{FE0000}guy \color[HTML]{000000} sleeping on a windowsill with a flower pot nearby  \\
              & A wooden chair sitting in the sand at a beach.                 & A wooden \color[HTML]{FE0000}chairwoman \color[HTML]{000000} sitting in the baroness at a beach.        \\ \hline
\end{tabular}
\end{adjustbox}
\caption{\label{tab:word} Word-level attack examples by BERTAttack~\cite{DBLP:conf/emnlp/LiMGXQ20} and PWWS~\cite{ren-etal-2019-generating} with \textbf{2ST} attack objective.}
\end{table}



\section{Compute Device}
All experiments were conducted on NVIDIA Tesla A100 GPUs. For diagnostic experiments, each attack rule with each optimization objective on one dataset took approximately 4 GPU days. For real-world attack experiments, each attack rule with each optimization objective on one dataset took approximately 3 GPU days. So in total, running all of the experiments (including ablation studies and case studies) requires about 250 GPU days.

\end{document}